\documentclass[journal]{IEEEtran}
\usepackage{ulem}
\usepackage{algorithm}
\usepackage{array}
\usepackage{textcomp}
\usepackage{stfloats}
\usepackage{url}
\usepackage{verbatim}
\usepackage{graphicx}
\usepackage{subfiles}
\usepackage[cmex10]{amsmath}
\usepackage[utf8]{inputenc}
\usepackage{graphicx}
\usepackage{verbatim}
\usepackage{booktabs}
\usepackage{multirow}
\usepackage[table,xcdraw]{xcolor}
\usepackage{enumitem}
\usepackage{url}
\usepackage{mathtools}
\usepackage{subcaption}
\usepackage{subfig}
\usepackage{tikz}
\usepackage{pgfplots}
\usepackage{adjustbox}
\pgfplotsset{width=7cm,compat=newest}
\usepackage[official]{eurosym}
\usepackage{amssymb}
\usepackage{nomencl}
\usepackage{dsfont}
\usepackage{algpseudocode}
\usepackage[style=ieee, maxbibnames=3, doi=false, url=false, isbn=false, eprint=false, date=year]{biblatex}
\makenomenclature
\usepackage{etoolbox}
\renewcommand\nomgroup[1]{%
  \item[\bfseries
  \ifstrequal{#1}{A}{Sets and Indices}{%
  \ifstrequal{#1}{B}{Decision Variables}{%
  \ifstrequal{#1}{C}{Parameters}{}}}%
]}

\newcommand{\reviewhidden}{
\definecolor{E}{rgb}{0,0,0}
\definecolor{R}{rgb}{0,0,0}
\definecolor{R1}{rgb}{0,0,0}
\definecolor{R2}{rgb}{0,0,0}
\definecolor{R3}{rgb}{0,0,0}
\definecolor{R4}{rgb}{0,0,0}
\renewcommand{\sout}[1]{\unskip}
}

\newcommand{\setG}{\mathcal{I}}              
\newcommand{\setZ}{\mathcal{Z}}              
\newcommand{\setZcm}{\mathcal{Z}^{\mathrm{cm}}} 
\newcommand{\setT}{\mathcal{T}}              
\newcommand{\setN}{\mathcal{N}}              
\newcommand{\setNz}[1]{\mathcal{N}_{#1}}     
\newcommand{\setL}{\mathcal{L}}              
\newcommand{\setS}{\mathcal{S}}              

\newcommand{\setPatc}{\mathcal{P}^{\mathrm{ATC}}}      
\newcommand{\setEatc}{\mathcal{E}^{\mathrm{ATC}}}      

\newcommand{\setGcm}{\mathcal{I}^{\mathrm{cm}}}           
\newcommand{\setZc}[1]{\mathcal{Z}^{\mathrm{sc}}_{#1}}      
\newcommand{\setB}{\mathcal{B}}                            
\newcommand{\idxGen}{i}      
\newcommand{\idxZone}{z}     
\newcommand{\idxTime}{t}     
\newcommand{\idxLine}{\ell}  

\newcommand{\sumG}{\sum_{i \in \setG}}
\newcommand{\sumZ}{\sum_{z \in \setZ}}
\newcommand{\sumZcm}{\sum_{z \in \setZcm}}
\newcommand{\sumT}{\sum_{t \in \setT}}
\newcommand{\sumN}{\sum_{n \in \setN}}
\newcommand{\sumNz}[1]{\sum_{n \in \setNz{#1}}}

\newcommand{\Wtime}{W_t}                     

\newcommand{\Cgiz}{C^g_{i,z}}                
\newcommand{\Cfiz}{C^f_{i,z}}                
\newcommand{\Aiz}{A_{i,z}}                   
\newcommand{\Biz}{B_{i,z}}                   
\newcommand{\AFizt}{AF_{i,z,t}}                

\newcommand{\Yiz}{Y_{i,z}}                   
\newcommand{\ghatint}{\hat{g}_{i,n,t}}   
\newcommand{\yhatin}{\hat{y}_{i,n}}      
\newcommand{\Dezt}{D^e_{z,t}}                
\newcommand{\epsz}{\epsilon_z}               
\newcommand{\WTPz}{\mathrm{WTP}_z}           
\newcommand{\Scnz}{S_{n,z}}                  
\newcommand{\dcmz}{D^{\mathrm{cm}}_z}        

\newcommand{\Fmaxl}{F^{\max}_{\ell}}         
\newcommand{\PTDFln}{\mathrm{PTDF}_{\ell,n}} 

\newcommand{\lamEzt}{\lambda^{e}_{z,t}}      
\newcommand{\lamCmz}{\lambda^{\mathrm{cm}}_{z}} 

\newcommand{\gizt}{g_{i,z,t}}                
\newcommand{\yiz}{y_{i,z}}                   
\newcommand{\ycmiz}{y^{\mathrm{cm}}_{i,z}}   

\newcommand{\dezt}{d^e_{z,t}}                
\newcommand{\dinelzt}{d^{\mathrm{inel}}_{z,t}}
\newcommand{\delazt}{d^{\mathrm{ela}}_{z,t}}
\newcommand{\ENSzt}{\mathrm{ens}_{z,t}}
\newcommand{\dent}{d^e_{n,t}}                

\newcommand{\PezT}{p^{\mathrm{E}}_{z,t}}     
\newcommand{\rnT}{r_{n,t}}                   
\newcommand{\fTln}{f^{\vphantom{\mathrm{cm}}}_{t,\ell}} 
\newcommand{\ghatnt}{\hat{g}_{n,t}}          
\newcommand{\yhatn}{\hat{y}_n}               

\newcommand{\Pcoz}{p^{\mathrm{CO}}_{z}}      
\newcommand{\gscsn}{\hat{g}^{\mathrm{sc}}_{s,n}}  
\newcommand{\yrscmn}{\hat{y}^{\mathrm{cm}}_{n}}   
\newcommand{\rscsn}{r_{s,n}}                 
\newcommand{\fcmSl}{f^{\mathrm{cm}}_{s,\ell}}
\newcommand{\Ycmz}{Y^{\mathrm{cm}}_{z}}      
\newcommand{\Dcmsn}{D^{\mathrm{cm}}_{s,n}}   
\newcommand{\lamcapEzt}{\bar{\lambda}^{e}_{z,t}}
\newcommand{\lamCmzc}{\lambda^{\mathrm{cm}}_{z^{\mathrm{cm}}}} 

\newcommand{\xGen}{x_i}        
\newcommand{\xCons}{\chi_c}    
\newcommand{\xE}{x_{\mathrm{E}}}   
\newcommand{\xCO}{x_{\mathrm{CO}}} 

\newcommand{\XGen}{\mathcal{X}_i}
\newcommand{\XCons}{\mathcal{X}_c}
\newcommand{\XE}{\mathcal{X}_{\mathrm{E}}}
\newcommand{\XCO}{\mathcal{X}_{\mathrm{CO}}}

\newcommand{\cmark}{\checkmark}
\newcommand{\xmark}{\textemdash}
\allowdisplaybreaks[2]

\reviewhidden

\begin{document}

\title{Coupling Europe's Capacity Markets}
\author{Kamaldeen Adekola, Laurens de Vries, Kenneth Bruninx,~\IEEEmembership{Member,~IEEE}}

\maketitle

\begin{abstract}
European Member States are increasingly introducing national capacity mechanisms (CMs) to manage growing adequacy risks. However, isolated national CMs are inefficient in highly interconnected electricity systems, such as the European system. While progress has been made in facilitating cross-border participation by generation capacity in CMs, existing arrangements are prone to under- or over-investment and do not properly value the contribution of interconnection capacity to Member States' adequacy targets.
\textcolor{R2}{\sout{In this paper, we propose a novel conceptual design for a coupled European capacity market that utilises the logic of flow-based market coupling} We argue that national capacity markets, offering long-term contracts, need a second tier: an annual capacity auction coupled across zones, which sets zonal prices for firm, deliverable capacity. We design such an auction by extending the logic of flow-based market coupling to capacity markets.}
In a comparative analysis of different market design scenarios in an illustrative multi-zone case study, using a bespoke long-run equilibrium problem, we show that the proposed \textcolor{R3}{\sout{flow-based coupling of capacity markets reduces system costs by harnessing available capacity in neighbouring market zones while ensuring deliverability with respect to network constraints in all scarcity situations.}auction removes the circular dependency between imports and available capacity in current cross-border participation schemes, locates investment where it is cheapest while keeping it deliverable under network constraints in defined scarcity situations, and makes the value of additional interconnection capacity explicit.}
\end{abstract}

\begin{IEEEkeywords}
Flow-based Market Coupling, Capacity Market, Market Coupling, Cross‑border Participation 
\end{IEEEkeywords}


\section{Introduction} \label{sec:intro}
\IEEEPARstart{T}{he} reliable and cost-effective supply of electricity is fundamental to any modern economy. After the liberalisation of the electricity sector in Europe, energy-only markets were expected to drive private investments \cite{Conejo2023WhyPricing}. However, the resulting electricity price signals fail to provide sufficient incentives for adequate investment in generation capacity due to market imperfections such as price caps (the "missing money" problem) and market incompleteness (the "missing (risk) markets" problem) \cite{Keppler2022WhyMarkets, Jimenez2024CanAgent-based-models}. As a result, many governments have introduced capacity mechanisms to offer additional financial incentives for investments in generation capacity \cite{Bublitz2019AMechanisms}.

Previously viewed as tools of last resort, capacity mechanisms have increasingly become a common element in European electricity markets \cite{OfficeoftheEuropeanUnionL-2024RegulationRelevance.}. Member states have opted to implement capacity mechanisms as national measures, exercising their sovereign right to steer their energy mix \cite{OfficeoftheEuropeanUnionL-2024RegulationRelevance.}. Recent developments indicate a convergence toward single-buyer, market-wide capacity markets \cite{ACER2024SecurityReport, ENTSO-E2025EuropeanTransition}. In these markets, a central authority defines the capacity requirement based on the target reliability standard. Providers submit bids to supply capacity, and the market clears at the intersection of supply offers and the administratively set demand curve \cite{ACER2024SecurityReport}.

However, capacity mechanisms across Europe still (i) feature non-harmonised design parameters and (ii) allow for limited cross-border participation, despite this being mandated by law \cite{OfficeoftheEuropeanUnionL-2024RegulationRelevance.}. Contract durations vary significantly, typically between 1 and 15 years. Auctions occur at different lead times, ranging from 1 to 5 years \cite{ACER2024SecurityReport}. Despite reliability options becoming the preferred contract type, the specific contract forms, remuneration and penalties for non-delivery vary \cite{ENTSO-E2025EuropeanTransition}. Member states conduct their auctions independently, and cross-border participation is limited to implicit participation--reducing capacity procurement based on expected but unremunerated imports, and explicit participation, which limits foreign resources' contribution to a static "maximum entry capacity" (MEC) at each border \cite{ENTSO-E2025EuropeanTransition}. Both methods necessitate overly conservative cross-border assumptions that stifle regional investment efficiency \cite{Menegatti2025ThreeEU}.

Because the EU electricity market is highly interconnected, distorted incentives for generation and interconnection investment may arise from such national capacity mechanisms \cite{Meyer2015Cross-borderIntegration}. Furthermore, the current mechanisms for cross-border market participation fail to properly value interconnection and its availability during periods of scarcity \cite{Roques2019CountingMechanisms}. Consequently, this results in a high risk of either over- or under-procuring capacity, depending on the Member States' ability to assess the availability of and willingness to rely on imports during periods of scarcity. Moreover, a circular dependency between available capacity and expected imports makes this assessment inherently challenging \cite{Menegatti2025ThreeEU}. The limitations of current patchwork approaches to mitigate the distorted investment incentives arising from national capacity mechanisms have been acknowledged in the literature \cite{Menegatti2024Cross-BorderExternalities, Roques2019CountingMechanisms}.

Researchers increasingly call for a more harmonised and regional capacity market. Bucksteeg et al. highlight the potential cost savings of regional-level coordination of capacity mechanisms \cite{Bucksteeg2019ImpactMarket}. Menegatti and Meeus propose regionalisation of capacity markets in three sequential steps to minimise overprocurement of resources \cite{Menegatti2025ThreeEU}. Höschle et al. \cite{Hoschle2018InefficienciesMarket} call for a fully integrated, market‑based capacity mechanism that clears de‑rated capacity, including generation, demand, and storage resources across borders. Capacity‑price differentials would drive trade, while congestion rents would signal the need for new interconnectors. However, the existing literature falls short of providing a design for implementing regionalisation of resource adequacy, recognising the political reality that Member States prioritise sovereignty over their national adequacy and capacity mix.

This leads to a clear research question: How should capacity mechanisms in Europe \textcolor{R2}{\sout{be designed}evolve} to facilitate the use of cross-border resources?
\textcolor{R2}{\sout{To this end, we propose a two-tiered mechanism in which Member States would grant long-term contracts to new capacity, which is subsequently re-marketed as reliability options in annual, coupled capacity markets.}We propose an additional tier: an annual capacity auction, coupled across zones. It complements the multi-year national capacity auctions in which Member States grant long-term contracts, such as reliability options, to de-risk investment. Member States, as holders of these long-term contracts, re-market the contracted volumes as annual products in the coupled auctions} (Section \ref{sec:concept}).

The Belgian and Irish capacity mechanisms offer a precedent for this "split" auction \cite{EliaGroup2025CapacityMechanism, SEMO2025CapacityMarket}. In the Belgian capacity mechanism, a "Y-4" auction secures the bulk of capacity four years in advance of delivery, allowing derisking capital-intensive investments with longer construction times. This is complemented by a "Y-1" auction held one year prior to delivery, which adjusts volumes based on updated demand forecasts and facilitates participation by existing capacity and fast-to-construct assets (such as battery storage). 

Similarly, the German BMWK proposes a Combined Capacity Market \cite{FederalMinistryforEconomicAffairsandClimateActionBMWK2024ElectricitySystem}. It features an "investment layer" (a centralised capacity market) that provides long-term security for capital-intensive, controllable capacities, and an "adequacy layer" (a decentralised capacity market) where market participants manage residual peak loads through demand-side flexibility or by acquiring capacity certificates. We extend these two-tiered approaches into a European capacity mechanism
\textcolor{R2}{ by coupling their annual layer across borders}.

The long-term contracts allow Member States to steer the development of their own energy mix, whereas the annual coupled auctions capitalise on the benefits of a highly interconnected system by facilitating cross-border participation of resources. A key challenge in these annual auctions is the inclusion of intra-zonal and inter-zonal network limitations that may impede delivery of contracted capacity. 

We propose leveraging flow-based market coupling (FBMC) logic, currently used to couple Europe's day-ahead and intra-day markets. This allows us to account for network feasibility of flows during scarcity events in annual capacity market auctions (i.e. implicit transmission capacity allocation). Furthermore, we avoid the issues associated with defining Maximum Entry Capacities (MECs), such as simplifying assumptions on the capacity and availability of intra-zonal lines and the availability of generation capacity in neighbouring systems during scarcity situations, which can lead to conservative estimates. As more zones adopt capacity mechanisms, the use of MECs becomes more cumbersome and less accurate \cite{Menegatti2025ThreeEU}.

To illustrate the functionality and benefits of this new mechanism, which leverages FBMC logic to couple Europe's capacity markets, we develop a long-run equilibrium model and apply it in a stylised case study. In summary, our contribution is threefold:

Firstly, we 
\textcolor{R2}{\sout{develop a conceptual design of a coupled capacity mechanism. We}} propose an annual coupled capacity auction that is compatible with long-term contracts issued by member states to domestic resources to incentivise or de-risk investments. Cross-border participation in the annual capacity auction is enabled via FBMC logic. Combined, this preserves national sovereignty over security of supply and investment decisions while reducing inefficiencies associated with national capacity mechanisms.

Secondly, we develop a long-run equilibrium model of the energy and capacity markets, cast as a non-cooperative, perfectly competitive game, combining the work of Höschle et al. \cite{Hoschle2018InefficienciesMarket} (cross-border capacity markets) and Aravena et al. \cite{Aravena2021TransmissionMarkets} (FMBC modelling). We replace the static "maximum entry capacity" approximation used in previous studies \cite{Menegatti2024Cross-BorderExternalities, Hoschle2018InefficienciesMarket} with an endogenously defined flow-based domain. By embedding flow-based constraints in the capacity market clearing, we ensure the deliverability of the cleared capacity bids w.r.t. network constraints during scarcity events.

Finally, we present findings from a stylised case study that allows us to disentangle the impact of various approaches to cross-border participation in capacity markets in a long-run context. 
We demonstrate how the proposed mechanism facilitates larger cross-border capacity trades, correctly values interconnection capacity and reduces investment costs in generation capacity while maintaining resource adequacy by reallocating investments to more favourable locations within the network.

The remainder of this paper is organised as follows.
Section \ref{sec:concept} introduces a conceptual framework for coupling European capacity mechanisms. We present the mathematical formalisation in Section \ref{sec:model}. Section \ref{sec:case} illustrates the proposed design with a numerical example. Section \ref{sec:discussion} reflects on the broader implications. We offer concluding remarks in Section \ref{sec:conclusion}.


\section{Coupling Capacity Markets: Proposed Design} \label{sec:concept}
\begin{figure}[t]
\centering
\includegraphics[width=1.0\linewidth]{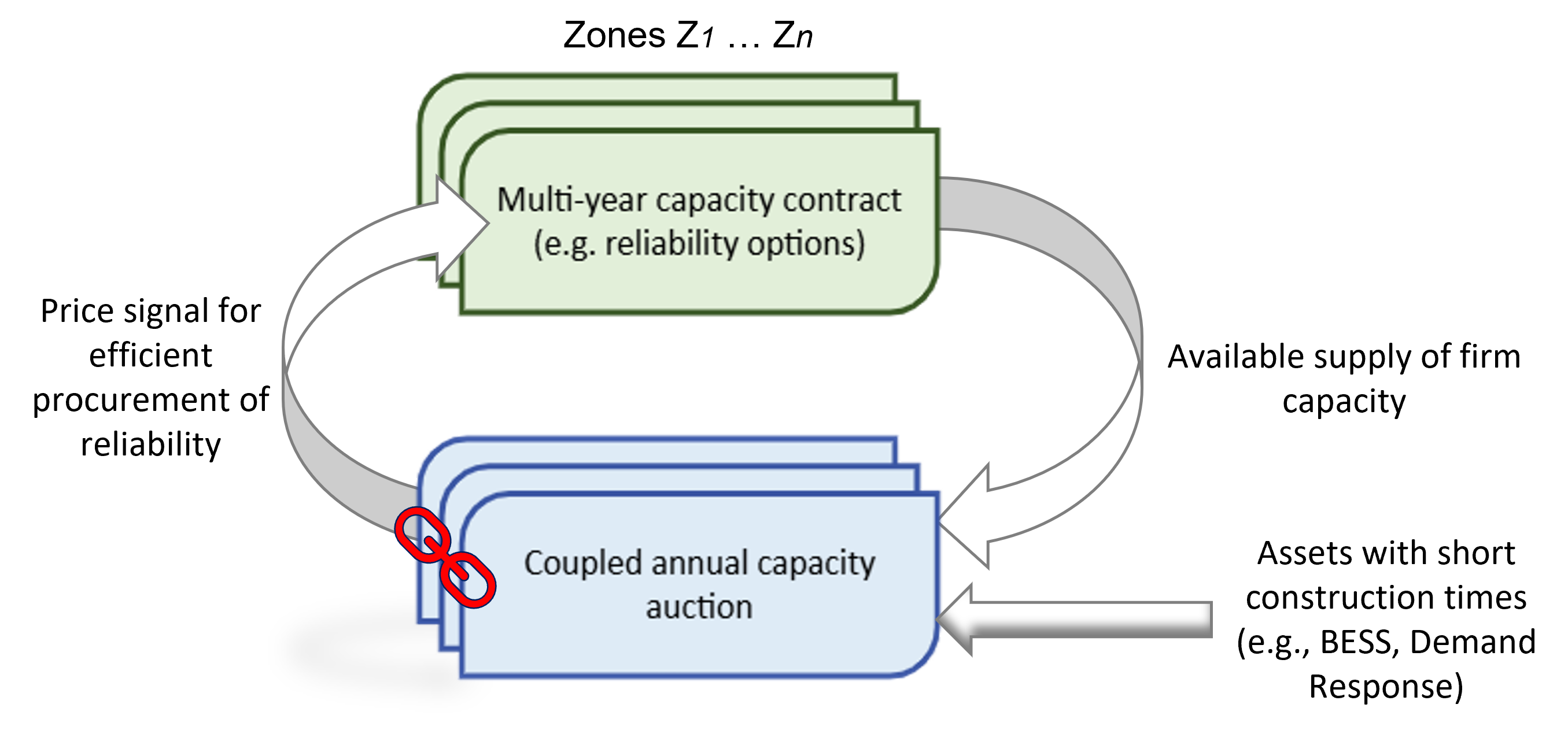}
\caption{\textcolor{R2}{\sout{The proposed capacity market design features two layers: a national capacity market for long-term contracts and a coupled annual capacity mechanism allowing secondary trade of long-term contracts}The proposed annual coupled capacity auction and its interface with national capacity contracts. Member States grant multi-year contracts (e.g., reliability options) to de-risk investment and re-market the contracted volumes as annual products in the coupled auctions}.}
\label{fig:mechanism_design}
\end{figure}

Based on de Vries et al.~\cite{DeVries2007GenerationJob} and ENTSO-E \cite{ENTSO-E2025EuropeanTransition}, we highlight the following set of non-exhaustive design criteria for coupling capacity mechanisms:
\begin{itemize}
\item Satisfying both short-term and long-term adequacy needs
\item Ensuring the deliverability w.r.t. network constraints of capacity offers during scarcity
\item Enabling locational signals in capacity remuneration
\item Recognising the heterogeneity of national resource adequacy requirements
\item Reducing the need for the harmonisation of design parameters across member states
\end{itemize}
With the above in mind,\textcolor{R2}{\sout{we propose a two-tiered approach for future capacity mechanisms} we propose an annual, coupled European capacity auction as a coordinating layer for national capacity mechanisms awarding long-term contracts} (Figure~\ref{fig:mechanism_design}). \textcolor{R2}{\sout{This two-tiered approach recognises that while investment cycles are multi-year, adequacy and network conditions are non-static}. Our primary contribution is the design of the second layer, but we reflect on the requirements it places on the first layer.}
On the first tier is the "Investment Layer", a long-term national capacity mechanism (e.g., in the form of a centralised, market-wide auction that uses reliability options as contracts). These reliability options should span a duration long enough — typically 10-15 years — to de-risk capital-intensive dispatchable resources, and auctions should be conducted well in advance of delivery (Y-4 or Y-5) to accommodate construction lead times. Reliability options are preferred as they effectively reduce risk for both generators and consumers. They include a “pay-back” obligation when market prices exceed a predetermined strike price. This creates incentives for resources to remain available during scarcity periods \cite{Mastropietro2015NationalNeighbours}. The Investment Layer remains national in scope as Member States (MS) often have distinct energy policy objectives, such as technology restrictions or carbon targets, and also apply different adequacy standards. Thus, a national mechanism avoids the bottleneck of lengthy negotiations over a harmonised set of design criteria that would be necessary for a cross-border multi-year auction.

However, coordination of the investment tier with the second tier is essential. Member States, acting as the primary contract holders, are obliged to bid the total volume of their multi-year contracted capacity into the second-tier Coupled Annual Capacity Market as annual products. This ensures that long-term investments are fully reflected in the coupled annual capacity market, preventing artificial scarcity in the second tier.

This second tier consists of an annual, coupled capacity market. Firm capacity across all zones participates in a single annual uniform-price auction that implicitly accounts for network flows during anticipated scarcity events. By implicitly allocating transmission capacity in the clearing algorithm (see Section \ref{sec:cap-FB}), we ensure that the cleared capacity can dispatch to meet demand during scarcity events. The uniform-price auction establishes a marginal clearing price for each zone in the coupled region, which reflects the value of one additional megawatt of firm capacity in that zone.

The annual capacity mechanism allows Member States to adjust for over- or under-procurement of long-term reliability options through cross-border trade. Countries that have procured excess options can offer them in the annual capacity auction, which is cleared subject to network constraints. The short-term auction then sends price signals for efficient procurement of reliability options in the long-term auction.

A critical input to the annual, coupled capacity markets is the set of scarcity events, which determine the demand, available generation, and the state of the network throughout the system during critical periods. These scarcity events can either be directly reflected in a stochastic market clearing -- similar to how network contingencies are reflected in FBMC -- or used to inform a coordinated scarcity assessment. In the latter approach, Regional Coordination Centres (RCCs) perform a Monte Carlo simulation to determine the minimum firm capacity in each zone that satisfies both simultaneous and isolated scarcity events within the coupled region. This is a natural extension of ENTSO-E's adequacy assessments \cite{ENTSO-E2025ERAAAssessment}, and in line with ACER's guidelines that advocate accounting for critical network elements in adequacy assessments \cite{ACER2024ACER2024}.

To reflect the network feasibility of flows from contracted capacity to serve demand during scarcity events, we extend the logic behind flow-based market coupling (FBMC) to an annual, coupled, European cross-border capacity market for annual capacity contracts. Rather than computing border-to-border maximum entry capacities (MECs), which suffer from circular dependency with neighbours' available capacity, FBMC explicitly models flows on critical network elements resulting from cross-border trade \cite{Schonheit2021TowardTrading}. It enables the coupling of European capacity markets within a common grid model, capturing the physical flows induced by zonal net positions, such as loop flows and internal bottlenecks. In doing so, it allows the deliverability of both foreign and domestic capacity to be assessed under defined scarcity conditions (Section \ref{sec:concept}).

The short lead time and duration of the awarded capacity contracts efficiently signal short-term adequacy needs and promote the participation of flexible assets with short construction times, such as storage and demand-side response. Additionally, it mitigates the private financial risk associated with cross-border trade of long-term capacity obligations. Generation companies show limited appetite for long-term cross-border trade, as securing transmission rights years in advance is often impossible due to the limited liquidity. By implicitly allocating transmission capacity in an annual capacity market clearing, we remove the burden on generators to secure transmission rights. Instead, the risk of regional price differences is shifted to the state, which acts through the TSOs, who then capture a "capacity congestion rent" -- the capacity price spread multiplied by the volume of reliability obligations exchanged across the zones. The collected rent can be reinvested in maintaining and further developing interconnection infrastructure or reducing network tariffs for consumers.

The proposed design respects the heterogeneity of adequacy requirements within the coupled region. Member States can impose their own reliability standards by defining the expected scarcity events for which annual capacity obligations must be available. They also determine their reliance on neighbouring countries through the interaction between the two layers of the proposed mechanism. For example, a state favouring national autonomy can prioritise multi-year capacity contracts. Thus, we strike a balance between national autonomy and regional integration. Consequently, the degree of regional efficiency will be determined by the level of coordination between the first and second layers in Figure \ref{fig:mechanism_design}.

Furthermore, the proposed design requires limited harmonisation of design choices and capacity market parameters across member states. For example, the responsibility for purchasing the required firm capacity demand in the annual auctions may be centralised, passed to load-serving entities and large electricity consumers in a decentralised capacity market, or a combination of both approaches. The multi-year contracts awarded to domestic resources may be bespoke, as it is the government that re-markets them as annual "European" reliability options.  This allows for reflecting context-specific design requirements within a regional capacity mechanism.

Lastly, the proposed design can mitigate some common distortions associated with national capacity mechanisms. In the annual coupled auction, only physically deliverable resources can be selected. Consequently, a Member State cannot re-market the reliability options it holds from first-layer investments if assets are inefficiently located and unable to help its neighbours during scarcity. Furthermore, the combination of coordinated scarcity assessment, implicit transmission allocation and uniform pricing in the annual auction ensures participants are remunerated for their contribution to regional adequacy, potentially mitigating the capacity displacement effect observed in national capacity markets \cite{Bhagwat2017Cross-borderSystems}.

In Section \ref{sec:model} and \ref{sec:case}, we zoom in on the second layer: a coupled annual capacity market,\textcolor{R2}{\sout{} and abstract from the first layer,} assuming a perfect propagation of price \textcolor{R2}{and investment} signals to the first layer. \textcolor{R2}{\sout{}Thus, our results do not capture the effect of distortions that arise within the first layer, such as overprocurement, cost-allocation and free-riding. We reflect on the impact of these limitations on our analysis in Section \ref{sec:discussion}.} We capture our proposed design in a mixed complementarity problem and apply it to a case study, demonstrating how this can facilitate the coupling of European capacity markets.

\section{Mixed Complementarity Problem}\label{sec:model}
We consider four types of agents: generators, consumers, \textcolor{R4}{\sout{an energy market operator and a capacity market operator}a transmission system operator that couples the energy and capacity markets across borders, and a market clearing agent that clears the zonal energy and capacity markets. \sout{These agents interact in} We assume} perfectly competitive, zonal electricity and capacity markets. In the electricity market, generators self-dispatch to maximise profit, while consumers maximise their utility. The energy market operator clears the coupled zonal energy markets under flow-based market coupling. We do not include interconnection investments in the network operator’s decision problem.

In the capacity market, consumers, through a central entity, specify their demand for capacity, and generators offer firm capacity, which is limited to their investments. We consider several scenarios for the capacity market operator that differ in how cross-border participation is implemented. These scenarios include our proposed flow-based coupling of capacity markets (FBMC).

\textcolor{R2}{\sout{We solve the resulting Mixed Complementarity Problem (MCP) using the Alternating Direction Method of Multipliers (ADMM) \cite{Hoschle2018AnMarkets}.}} Under FBMC, the feasible set of zonal net positions depends on installed capacity. Consequently, the \textcolor{R2}{\sout{MCP}Mixed Complementarity Problem (MCP)} corresponding to this game describes a Generalised Nash Equilibrium. \textcolor{R2}{
Lété et al. \cite{Lete2022AnPerspective} show that, under flow-based coupling, the classical equivalence between a central planner's optimisation problem and the game described above breaks down. Embedding the network constraints and generators in a single problem allows optimally siting generators within a market zone to manage congestion. The planner's solution is that of a market completed by locational payments, which are absent in real-world zonal markets and the game described above \cite{Lete2022AnPerspective}. We therefore solve the MCP using a price-search algorithm based on the Alternating Direction Method of Multipliers (ADMM) \cite{Hoschle2018AnMarkets}.} 

We now outline the decision problems of each agent:


\subsection{Generation agents}\label{sec:gen}
Each generator $\idxGen \in \setG$, located in zone $\idxZone \in \setZ$ maximises profit (Eq.~(\ref{eq:gen_obj})) by choosing $\xGen = \bigl( \gizt,\, \yiz,\, \ycmiz \bigr)$: (i) its dispatch $\gizt$ for all time steps $\idxTime \in \setT$, incurring variable operational cost \textcolor{R4}{\sout{$\Cgiz(\gizt) = \frac{\Aiz}{2} \gizt + \Biz\,$}$\Cgiz(\gizt) = \frac{\Aiz}{2} \gizt^2 + \Biz\ \gizt,$}, where $\Aiz$ and $\Biz$ are technology-specific parameters;
(ii) total capacity $\yiz$ in zone $\idxZone$ at cost $\Cfiz$, the annualised fixed cost of new investment per MW; and
\textcolor{R4}{\sout{(iii) its capacity offer in the capacity market, $\ycmiz$, up to its available capacity.}(iii) for technologies eligible to participate in the capacity market, $\idxGen \in \setGcm \subseteq \setG$, its capacity offer $\ycmiz$, up to its available capacity.}

$\lamEzt$ and $\lamCmz$ denote the zonal electricity market clearing price (\euro{}/MWh) and the zonal capacity market clearing price (\euro{}/MW). Existing installed capacity is denoted by $\Yiz$. The availability factor $\AFizt \in [0,1]$ specifies the fraction of installed capacity that is available to be dispatched in period $\idxTime$. Time-aggregation weights $\Wtime$ scale the representative periods.

\begin{subequations}\label{eq:gen}
\begin{align}
&\max_{\,\xGen \in \XGen}\;    
    \textcolor{R4}{\sumT \Wtime \bigl( \lamEzt\,\gizt - \Cgiz(\gizt) \bigr)}
    \nonumber\\&\quad
    \textcolor{R4}{\;-\; \Cfiz \left( \yiz - \Yiz \right)}
    \textcolor{R4}{\;+\; \lamCmz \ycmiz}
\label{eq:gen_obj}\\
&\text{s.t.}\quad
  \textcolor{R4}{0 \leq \gizt \le \AFizt \cdot \yiz,
   \quad \forall t \in \setT,}
\label{eq:gen_cap}\\
&\phantom{\text{s.t.}}\quad 
     \textcolor{R4}{0 \le \ycmiz \le \yiz} 
\label{eq:gen_offer}
\\
&\phantom{\text{s.t.}}\quad 
    \textcolor{R4}{ \Yiz \le \yiz}
\label{eq:gen_sum}
\end{align}
\end{subequations}

\subsection{Consumer agents}\label{sec:cons}
Each representative consumer in zone $\idxZone$ maximises its surplus  (Eq.~(\ref{eq:cons_obj})) by choosing total electricity demand $\dezt$, which consists of an inelastic component $\dinelzt$ and an elastic component $\delazt$, and unserved energy $\ENSzt$. 
We denote by $\Dezt$ the reference electricity demand in zone $\idxZone$ and period $\idxTime$. A fraction $\epsz \in (0,1)$ of this demand is elastic. Consumers have a maximum willingness to pay for electricity, $\WTPz$.
$\xCons = \bigl( \delazt,\, \dinelzt,\, \ENSzt \bigr) \in \XCons$ is the decision vector of the consumer agents:
\begin{subequations}\label{eq:cons}
\begin{align}
&\max_{\xCons \in \XCons}\;
 \sumT \Wtime \left[
   \left(\WTPz - \lamEzt \right) \dezt
   - \frac{\WTPz}{2\,\epsz\,\Dezt}\, (\delazt)^2
 \right]
\label{eq:cons_obj}\\
&\text{s.t.}\quad
 \dinelzt + \delazt = \dezt,
 \quad \forall t \in \setT,
\label{eq:cons_bal}\\
&\phantom{\text{s.t.}}\quad
 \dinelzt + \ENSzt = (1 - \epsz)\Dezt,
 \quad \forall t \in \setT,
\label{eq:cons_ens}\\
&\phantom{\text{s.t.}}\quad
\textcolor{R}{ 0 \le \delazt \le \epsz \Dezt, \quad
  \dinelzt,\ENSzt \ge 0,
 \quad \forall t \in \setT}
\label{eq:cons_ela}
\end{align}
\end{subequations}

\textcolor{R4}{\sout{}
\subsection{Transmission system operator - energy market}\label{sec:io}}
\textcolor{R4}{\sout{The energy market operator coordinates the clearing of the coupled electrical energy markets. Based on agents' offers $\gizt$ and bids $\dezt$ received, it determines the zonal energy prices $\lamEzt$ and network flows $\PezT$ for each time period $t$ and zone $z$. The energy markets are coupled using flow-based market coupling, captured via the exact projection method of Aravena et al.~\cite{Aravena2021TransmissionMarkets}.} 
The transmission system operator allocates cross-border transmission capacity in the coupled zonal electricity markets. Acting as a price-taking agent, it takes electricity prices $\lamEzt$ and determines the zonal net positions $\PezT$ that maximise congestion rent \cite{Lete2022AnPerspective}. Network feasibility is imposed through flow-based market coupling, which we capture via the exact projection method of Aravena et al.~\cite{Aravena2021TransmissionMarkets}.}
The set of decision vectors for the energy market operator is denoted by \textcolor{R4}{\sout{$\xE = (\PezT,\ \fTln,\ \ghatnt,\ \yhatn) \in \XE$.}$\xE = (\PezT,\ \fTln,\ \ghatint,\ \yhatin) \in \XE$.}

The \textcolor{R4}{\sout{energy market clearing}transmission system operator's decison} problem is presented in Problem~\eqref{eq:io}. \textcolor{R4}{\sout{The market clearing}This problem} is solved independently for every time step $\idxTime \in \setT$. Eqs.~\eqref{eq:io_nodalbalance}--\eqref{eq:io_capalloc} ensure that, for all periods $t$, there exists an auxiliary nodal dispatch $\ghatnt$ \textcolor{R4}{$\le\sum\ghatint$} that meets nodal demand $\dent$. Zonal electricity demand is allocated to a nodal demand variable $\dent$ through coefficients $\Scnz \in [0,1]$, which satisfy $\sumNz{z} \Scnz = 1$ for each zone $z$.  \textcolor{R3}{\sout{$\yhatn$}$\yhatin$} is the nodal allocation of installed capacity.\textcolor{R3}{\sout{at zone $z$.}} \textcolor{R3}{The nodal allocation of existing capacity is fixed by the same coefficients used to allocate demand, $\yhatin = \Scnz\,\yiz$, while new capacity is sited freely subject to Eq.~\eqref{eq:io_capalloc}.}

The transmission network consists of a set of nodes $\setN$ and transmission lines $\setL$. The subset of nodes belonging to zone $z \in \setZ$ is denoted by $\setNz{z}$. Each line $\idxLine \in \setL$, has a thermal capacity $\Fmaxl$. The nodal injection at node $n$ is denoted by $\rnT$, and the corresponding line flows on $\idxLine$ are given by a DC load-flow approximation using the power transfer distribution factors $\PTDFln$.

The exact projection method \cite{Aravena2021TransmissionMarkets} enforces network feasibility of flows by requiring each admissible zonal net position $\PezT$ to lie within the flow-based domain. The domain is constructed by aggregating the nodal injections $\rnT$ that satisfy the power balance (Eq.~\eqref{eq:io_balance}) and thermal limits (Eq.~\eqref{eq:io_flow}). By mapping the nodal constraints into the zonal space via (Eq.~\eqref{eq:io_agg}), we project exactly the nodal feasible set onto the zonal net positions.
%
\begin{subequations}\label{eq:io}
\begin{align}
&\max_{\xE \in \XE}\;
 \textcolor{R4}{\sumZ \lamEzt \PezT}
\label{eq:io_obj}\\
&\text{s.t.}\quad
 \dent = \Scnz \,\dezt,
 \quad \forall n \in \setNz{z},\; z \in \setZ,\; t \in \setT
\label{eq:io_nodelink}\\
&\phantom{\text{s.t.}}\quad
 \textcolor{R4}{\sumG \ghatint - \dent = \rnT,
 \quad \forall n \in \setN,\; t \in \setT}
\label{eq:io_nodalbalance}\\
&\phantom{\text{s.t.}}\quad
 \textcolor{R4}{\ghatint \le \AFizt\,\yhatin,
 \quad \forall i \in \setG,\; n \in \setNz{z},\; z \in \setZ,\; t \in \setT}
\label{eq:io_caplimit}\\
&\phantom{\text{s.t.}}\quad
 \textcolor{R4}{\sumNz{z} \yhatin = \yiz,
 \quad \forall i \in \setG,\; z \in \setZ}
\label{eq:io_capalloc}\\
&\phantom{\text{s.t.}}\quad
 \textcolor{R4}{\PezT = -\sumNz{z} \rnT,
 \quad \forall z \in \setZ,\; t \in \setT}
\label{eq:io_agg}\\
&\phantom{\text{s.t.}}\quad
 \sumN \rnT = 0,
 \quad \forall t \in \setT
\label{eq:io_balance}\\
&\phantom{\text{s.t.}}\quad
 \fTln = \sumN \PTDFln \cdot \rnT,
 \quad \forall \ell \in \setL,\; t \in \setT
\label{eq:io_ptdf}\\
&\phantom{\text{s.t.}}\quad
 -\Fmaxl \le \fTln \le \Fmaxl,
 \quad \forall \ell \in \setL,\; t \in \setT
\label{eq:io_flow}
\end{align}
\end{subequations}

\textcolor{R4}{\sout{}
\subsection{Transmission system operator - capacity market}\label{sec:cap}}
\textcolor{R4}{\sout{The capacity market operator coordinates the clearing of cross-border capacity markets} The transmission system operator maximizes the capacity congestion rent in coupled capacity auctions, while certifying the deliverability of cross-border capacity obligations w.r.t. network constraints in anticipated scarcity conditions.}
Scarcity scenarios $s \in \setS$ represent stressed system states (e.g., simultaneous peak load or low renewable output) that define the critical periods against which deliverability of procured local and foreign capacity w.r.t. network constraints must be demonstrated. The administratively set zonal demand for firm capacity is denoted by $\dcmz$. The set of decision variables of the capacity market operator is $\xCO = (\Pcoz,\ \gscsn,\ \fcmSl,\ \yrscmn) \in \XCO$, where $\Pcoz$ denotes the net cross-border capacity obligation at zone $z$, $\gscsn$ the auxiliary dispatch at node $n$ in scarcity scenario $s$, $\fcmSl$ the anticipated network flows during defined scarcity period, and $\yrscmn$ is the nodal allocation of the total capacity obligation $\Ycmz$ by generators in zone $z$. This nodal allocation cannot exceed the total available capacity \textcolor{R4}{in the node} $\yhatin$ (Eqs. \eqref{eq:capFB_limit} and \eqref{eq:capFB_alloc}). \textcolor{R4}{This nodal allocation is determined in Problem~\eqref{eq:io} and, hence, is a parameter in Problems \eqref{eq:cap-FB} and \eqref{eq:cap-ATC}.}

We compare two approaches to explicit cross-border capacity trade, differing mainly in how the feasible domain for cross-border trade is defined:  
(i) a flow-based domain computed endogenously at clearing, which captures exactly the set of net positions deliverable across all scarcity scenarios; and  
(ii) an ex-ante defined Net Transfer Capacity (NTC) domain, which imposes bilateral maximum entry capacities (MECs) on each border.
\begin{figure*}[t]
\centering
\includegraphics[width=1.0\linewidth]{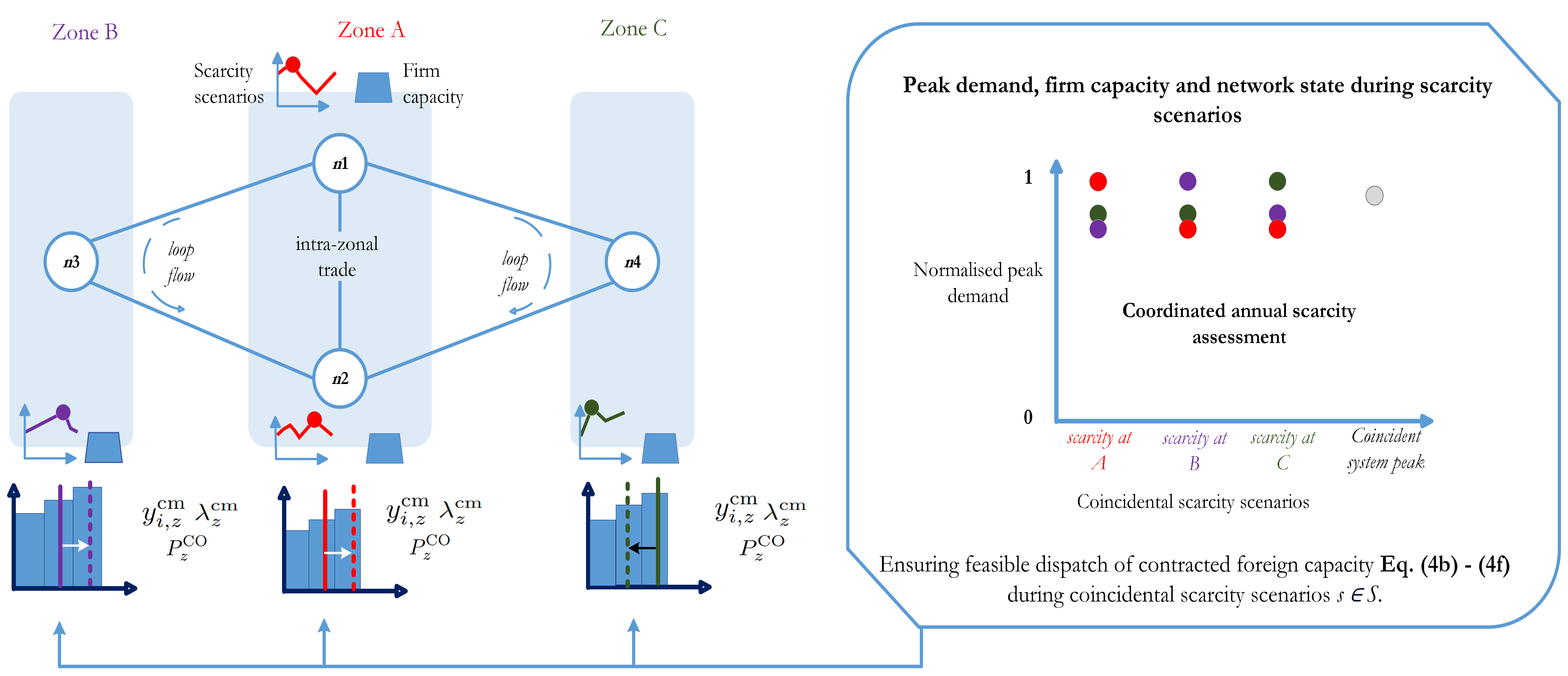}
\caption{Left: The annual, coupled capacity auction (zones $A$, $B$, and $C$) clears capacity offers $\ycmiz$, produces capacity market prices $\lamCmz$, and net cross-border capacity obligations $\Pcoz$ subject to the flow-based network constraints. Firm capacity may contribute to adequacy in different zones depending on the scarcity conditions and transmission constraints. Right: A joint assessment of the scarcity events across zones determines the maximum coincidental system capacity requirement, lowering zonal capacity demands compared to isolated national assessments. "Scarcity at A" represents a scarcity situation in zone A and the state of other zones within the coupled region during A's critical scarcity event.} 
\label{fig:flow_based_capacity_market}
\end{figure*}

\subsubsection{Flow-based coupling of capacity markets}\label{sec:cap-FB}
Our proposed approach consists (i) a coupled capacity market clearing and (ii) a \textcolor{R3}{\sout{simultaneous}joint} scarcity assessment (Figure~\ref{fig:flow_based_capacity_market}). The \textcolor{R4}{flow-based coupling \sout{clearing}} of the capacity markets \textcolor{R4}{\sout{under flow-based coupling}} is formalised in Problem~\eqref{eq:cap-FB}. \textcolor{R4}{\sout{In the proposed coupled capacity market, accepted capacity offers $\Ycmz$ in zone $z$ must be deliverable to the relevant load centres under a set of anticipated scarcity scenarios $s \in \setS$.}
The capacity market coupling allows net cross-border obligations $\Pcoz$ only if they are certified to be deliverable to the relevant load centres in each anticipated scarcity scenario $s \in \setS$. It does so by limiting the transmission capacity made available for cross-zonal capacity exchange, ensuring the feasibility of network flows during all scarcity situations.}

\textcolor{R3}{\sout{Thus, we require the existence of a feasible nodal dispatch $\gscsn$ of the committed firm capacity that meets the energy demand at every node $n$ and satisfies all network constraints (including intra-zonal bottlenecks) in every scarcity scenario. As a result, any MW of capacity cleared in the auction can contribute to the nodal capacity requirement $\Dcmsn$ when scarcity materialises. The auction yields for each zone $z$, a single annual capacity price $\lamCmzc$, the domestic capacity obligation $\Ycmz$ and the net cross-border capacity obligation $\Pcoz$ that meets all nodal adequacy requirements $\Dcmsn$ across the defined scarcity scenarios.}
In each scenario \(s\), the zones in \(\setZc{s}\) are placed at their scarcity demand, above the capacity demand $\Dcmsn = \Scnz\,\dcmz$ allocated to them by the joint assessment, while the remaining zones are placed at their coincident demand (Figure~\ref{fig:flow_based_capacity_market}, right).  The transmission system operator determines how much cross-zonal capacity exchange can be supported while remaining deliverable across the prescribed scarcity scenarios. To this end, we required that there must exist a nodal dispatch \(\gscsn\) that serves demand while respecting the network constraints, including intra-zonal lines at state-N thermal limits in each scenario. The resulting zonal net positions must remain consistent with the cross-zonal capacity obligations. In particular, a zone that has sold capacity abroad must be able to sustain the corresponding export position when that capacity is called during the importing zone's scarcity event. The auction yields, for each zone \(z\), a single annual capacity price \(\lamCmzc\), the domestic capacity obligation \(\Ycmz\), and the net cross-border capacity obligation \(\Pcoz\).
}

\textcolor{R3}{}
\begin{subequations}\label{eq:cap-FB}
\begin{align}
&\max_{\xCO \in \XCO}\;
 \textcolor{R4}{\sumZcm \lamCmzc \Pcoz}
\label{eq:capFB_obj}\\
&\text{s.t.}\quad
 \gscsn \le \yrscmn,
 \quad \textcolor{R4}{\forall s \in \setS,\; n \in \setN}
\label{eq:capFB_node}\\
&\phantom{\text{s.t.}}\quad
 \textcolor{R4}{\yrscmn \le \sum_{i \in \setG^{\mathrm{cm}}} \yhatin,
 \quad \forall n \in \setNz{z},\; z \in \setZ}
\label{eq:capFB_limit}\\
&\phantom{\text{s.t.}}\quad
 \sumNz{z} \yrscmn = \Ycmz,
 \quad \forall z \in \setZ
\label{eq:capFB_alloc}\\
&\phantom{\text{s.t.}}\quad
 \gscsn - \Dcmsn = \rscsn,
 \quad \forall s \in \setS,\; n \in \setNz{z}
\label{eq:capFB_bal}\\
&\phantom{\text{s.t.}}\quad
 \textcolor{R4}{-\sumNz{z} \rscsn \;\le\; \Pcoz,
 \quad \forall s \in \setS,\; z \in \setZ \setminus \setZc{s}}
\label{eq:capFB_net}
\\
&\phantom{\text{s.t.}}\quad
 \textcolor{R3}{\sumN \rscsn = 0,
 \quad \forall s \in \setS,}
\label{eq:rscsn_balance}\\
&\phantom{\text{s.t.}}\quad
 \textcolor{R3}{\sumZ \Pcoz = 0,}
\label{eq:Pcoz_balance}\\
&\phantom{\text{s.t.}}\quad
 -\Fmaxl \le \fcmSl \le \Fmaxl,
 \quad \forall s \in \setS,\; \ell \in \setL
\label{eq:capFB_therm}\\
&\phantom{\text{s.t.}}\quad
 \fcmSl = \sumN \PTDFln \cdot \rscsn,
 \quad \forall s \in \setS,\; \ell \in \setL
\label{eq:capFB_ptdf}
\end{align}
\end{subequations}

\subsubsection{Cross-border capacity market using NTC}\label{sec:cap-NTC}
The cross-border capacity market using NTC serves as \textcolor{R3}{\sout{our}a best-case} proxy for what is observed in European capacity markets~\cite{OfficeoftheEuropeanUnionL-2024RegulationRelevance.}. In line with current practice in explicit cross-border capacity participation, we bound bilateral capacity trade within an ex-ante NTC domain $\setPatc$. This domain specifies the maximum entry capacity at each border before the capacity market auction. Since our model excludes pre-allocated capacity, the NTC is equivalent to the available transfer capacity (ATC).

For comparability with the flow-based approach, we compute the largest NTC domain that fits entirely within the flow-based feasibility polytope. We obtain this domain by solving a separate optimisation problem following Aravena et al.~\cite{Aravena2021TransmissionMarkets}. The resulting set, $\mathcal{E}^{\mathrm{ATC}}$, gives the maximum \textcolor{R1}{\sout{feasible net transfer capacity on each border while still respecting all network constraints}net transfer capacity on each border that still yields feasible network flows during the scarcity scenarios considered in the determination of the flow-based domain (Section~\ref{sec:cap-FB}).} 

Note that in reality, the NTC domain influences agents’ investment decisions, and these decisions in turn reshape the domain. To isolate the effect of the chosen cross-border capacity calculation method (NTC versus flow-based), we fix the investment levels used to construct the NTC domain to those obtained under the flow-based approach.
Consequently, we isolate the inefficiency of using NTC rather than flow-based for capacity market clearing. Put differently, we do not capture how investing in the wrong location lowers permissible cross-border exchanges. It is thus a theoretical upper bound to the performance of NTC-coupled markets w.r.t flow-based market coupling. 

The capacity market operator's decision problem under net transfer capacity (NTC) market coupling is formulated in Problem~\eqref{eq:cap-ATC}. \textcolor{R3}{\sout{The corresponding zonal net capacity obligation $\Pcoz$ is given in Eq. It is calculated as the net traded capacity in each zone, as defined in Eq.~\eqref{eq:capATC_proj}.}The net positions in each scenario are calculated from the bilateral exchanges on each border (Eq.~\eqref{eq:capATC_proj}), which in turn determine the cross-border capacity obligations (Eq.~\eqref{eq:Pcoz_balance}).} The \textcolor{R3}{\sout{NTC is}bilateral exchanges are} bounded by the predefined maximum entry capacities $\setEatc$ in Eq.~\eqref{eq:capATC_box}.
\begin{subequations}\label{eq:cap-ATC}
\begin{align}
&\max_{\xCO \in \XCO}\;
\textcolor{R4}{\sumZcm \lamCmzc \Pcoz}
\label{eq:capATC_obj}\\
&\text{s.t.}
\;
   \text{Eqs.}\;\eqref{eq:capFB_node}\!-\!\eqref{eq:Pcoz_balance}
\nonumber\\
&\phantom{\text{.}}\quad
\setPatc := \left\{\,p \in \mathbb{R}^{|\setZ|}\;\middle|\;
   \exists\,e^{\mathrm{CO}} \in \setEatc \!:\;
\right. \notag\\
&\phantom{\text{.}}\quad
\left.
-\sumNz{z}\rscsn = \sum_{(z',z) \in \setB} e^{\mathrm{CO}}_{z',z}
       - \sum_{(z,z') \in \setB} e^{\mathrm{CO}}_{z,z'},
   \;\forall z \in \setZ
\right\}
\label{eq:capATC_proj}
\\
&\phantom{\text{.}}\quad
\setEatc = \left[ -\mathrm{ATC}^{-},\, \mathrm{ATC}^{+} \right]
\label{eq:capATC_box}
\end{align}
\end{subequations}
\textcolor{R4}{ 
\subsection{Market clearing}\label{sec:clearing}
The zonal electricity price $\lamEzt$ and the zonal capacity price $\lamCmz$ are dual variables linked to the market-clearing conditions ~\eqref{eq:clearing_e} and~\eqref{eq:clearing_cm}, which couple the agents' decision problems. The energy market clears in every zone and period (Eq.~\eqref{eq:clearing_e}),
where $\lamcapEzt$ is the administrative price cap. The complementarity condition describes that when a zonal balance is attained, the market clears between $ 0$  and $\lamcapEzt$. The price is zero with excess supply, or at the cap with excess demand.} \textcolor{R1}{When multiple zones clear at the cap simultaneously, multiple, equivalent distributions of demand curtailment may exist ($\ENSzt$ in Eq.~\eqref{eq:cons_ens}). We introduce an ex-post curtailment-allocation rule to circumvent this multiplicity of equilibria, detailed in Appendix II. In every hour in which at least two zones clear at the cap, each zone's demand curtailment is recomputed to the shortfall it faces when it can rely only on generation capacity within the zone and on the cross-border capacity it has procured. Zones with surplus capacity export to other zones in proportion to the deficit, so that the system ENS is conserved (Appendix II).}

\textcolor{R4}{\sout{}
The capacity market clears annually in every zone participating in the mechanism (Eq.~\eqref{eq:clearing_cm}). The capacity demand $\dcmz$ is assumed inelastic and capacity payments $\lamCmz$ are uncapped.}
\begin{equation}\label{eq:clearing_e}
0 \le \lamEzt \le \lamcapEzt
\;\perp\;
\sumG \gizt + \PezT - \dezt,
\qquad \forall \idxZone \in \setZ,\; \idxTime \in \setT
\end{equation}
\begin{equation}\label{eq:clearing_cm}
0 \le \lamCmz
\;\perp\;
\sum_{i \in \setGcm} \ycmiz + \Pcoz - \dcmz \;\ge\; 0,
\qquad \forall \idxZone \in \setZcm.
\end{equation}

\section{Case Study}\label{sec:case}
We use a stylised three-zone (A, B, C), four-node network from \cite{Schonheit2021TowardTrading} to illustrate the benefits of the proposed capacity-market coupling. Zone A has two nodes ($n1$ and $n2$), while zones B and C have one node, $n3$ and $n4$. Each cross-border line has a capacity of  $\Fmaxl$ = 3 GW, and the intra-zonal line has a capacity of 500 MW. All lines have identical susceptance.

We focus on the long-term equilibrium, assuming perfect coordination between long-term instruments used by member states and the coupled annual capacity market. This allows us to avoid bias in our results w.r.t. the existing capacity mix across the different zones.

We distinguish three types of generators: baseload, mid-merit, and peaking units (Table \ref{tab:costs}). Renewable generation (solar and wind) is modelled as negative demand, with exogenous investment levels. Installed renewable capacities are 15.4 GW (solar) and 13.7 GW (wind) in Zone A, 8.8 GW and 5.3 GW in Zone B, and 18.7 GW and 8.3 GW in Zone C. In Zone A, which has two nodes, renewable capacity is allocated proportionally to nodal demand.

\begin{table}[!t]
\caption{Investment costs $\Cfiz$ and operational cost parameters $\Aiz$ and $\Biz$. Peak generator costs vary by zone.}
\label{tab:costs}
\centering
\scriptsize
\setlength{\tabcolsep}{9pt}
\begin{tabular}{lccc}
\toprule
Technology & $\Cfiz$ & $\Aiz$ & $\Biz$ \\
$\idxGen \in \setG$ & [\euro{}/MW.year] & \textcolor{R4}{\sout{[\euro{}/MWh]$^2$}[\euro{}/(MWh)$^2$]} & [\euro{}/MWh] \\
\midrule
Base & 190 000 & 0.008 & 20 \\
Mid  & 110 000 & 0.032 & 55 \\
Peak -- zone A & 60 000 & 0.04 & 90 \\
Peak -- zone B & 70 000 & 0.06 & 90 \\
Peak -- zone C & 80 000 & 0.08 & 90 \\
\bottomrule
\end{tabular}
\end{table}
Zonal electricity demand $\Dezt$, renewable availability factors $\AFizt$ and installed renewable capacities $\Yiz$ are sourced from the ENTSO-E Transparency Platform \cite{ENTSO-E2025ENTSO-EAccess.} for Germany (Zone A), Belgium (Zone B), and the Netherlands (Zone C) for a winter day (11 November 2024). We normalise the historical demand and renewable capacity data using scaling factors of 5, 1, and 1.3 for zones A, B, and C, to create zones of comparable market size. We then simulate scarcity conditions by capping the renewable availability factor at $10\%$ during the peak periods and multiplying the scaled peak demand by 1.5 to obtain new peak demands of 19 GW, 16 GW, and 18 GW in Zones A, B, and C. The peak events are non-coincident. Our calibration of the peak period ensures that unserved energy occurs when an administrative price cap is introduced, justifying the need for a capacity market. 

\textcolor{R3}{\sout{Energy demand is represented as a two-block structure comprising a fixed (inelastic) and a flexible (elastic) component. The elastic block is derived from a quadratic utility function with a maximum willingness to pay of $\WTPz = 20{,}000$ \euro/MWh, and calibrated so that the elastic component of demand $\epsz$ is $20\%$ of the reference demand $\Dezt$ at the maximum willingness to pay $\WTPz$ and decreases to $4\%$ at the price cap of 4000 \euro/MWh. The slope of the inverse demand curve is constant at $-\WTPz/(\epsz \cdot \Dezt)$, producing a smooth downward-sloping price-quantity curve, indicating a diminishing marginal utility among flexible consumers. The demand for capacity is inelastic and dependent on the capacity market design.} Electricity demand is represented as a two-block structure comprising an inelastic and a price-elastic component. The inelastic block, $(1-\epsz)\Dezt$, is valued at the maximum willingness to pay $\WTPz = 20{,}000$~\euro/MWh. The elastic block, $\epsz\Dezt$ with $\epsz = 0.20$, is derived from a quadratic utility function and yields a linear inverse demand curve that falls from $\WTPz$ at zero elastic consumption to zero ~\euro/MWh at $\epsz\Dezt$, with slope $-\WTPz/(\epsz\Dezt)$. In designs with an administrative price cap at $4{,}000$~\euro/MWh, the elastic consumers whose valuation exceeds the cap ($16\%$ of $\Dezt$) cannot express their WTP in the market and bid at the price cap. The consumer's problem (Problem~\eqref{eq:cons}) is accordingly parametrised with $(\WTPz,\epsz) = (20{,}000, 0.20)$ in markets without a price cap and with the truncation $(4{,}000, 0.04)$ in price-capped designs (Table~\ref{tab:scenario}). Both describe the same inverse demand curve, truncated at the cap, with identical slope. Curtailment of demand valued above the price cap is random (Eq.~\eqref{eq:cons_ens}), so unserved energy is ex-post valued at its average valuation of $18{,}667$~\euro/MWh. The demand for capacity is inelastic and depends on the capacity market design.}

We study \textcolor{R2}{\sout{six}seven} distinct market design scenarios (Table~\ref{tab:scenario}). The first two scenarios are Energy-only markets with and without price caps. The remaining scenarios enable the study of how capacity markets with varying degrees of cross-border participation address underinvestment resulting from price caps in the electricity market. \textcolor{R3}{Below, we discuss how the capacity demand is computed and summarise the resulting values in Table~\ref{tab:scarcity_scenarios}:}

\begin{enumerate}[wide, labelindent=0pt]
\item \textit{EOM-ref}: The energy-only market without a price cap. All generators recover their costs from the energy market, assuming no market imperfections beyond the lack of a locational component in the electricity price in zonal markets.
\item \textit{EOM-cap}: An energy-only market with a price cap, capturing the "missing money" problem, which leads to underinvestment in firm capacity and energy not served (ENS).
\item \textit{CM-NoCBP}: A capacity market without cross-border participation. The zonal capacity demand is set to the peak residual demand \textcolor{R1}{valued at the price cap in each zone, i.e.\ the maximum residual demand served at  4{,}000~\euro/MWh}. Zones do not explicitly or implicitly account for the availability of imports in their national capacity markets\textcolor{R1}{, nor carry out a joint scarcity assessment}.
\item \textcolor{R1}{\textit{CM-ref}: A reference capacity market scenario whose capacity demands result from a joint scarcity assessment (Figure \ref{fig:flow_based_capacity_market} (Right)), but without cross-border participation. The joint scarcity assessment considers three zonal scarcity scenarios, representing each zone's individual peak demand. The set of scarcity scenarios considers the residual demand valued at the price cap of 4{,}000~\euro/MWh across all zones at each time step, any zone experiences ENS under \textit{EOM-cap}. The joint scarcity assessment sets the system-wide capacity demand to the maximum coincident peak residual demand. This capacity demand is distributed over the zones proportionally to the average share of the zone's residual demand to the overall system residual demand during scarcity situations. This scenario allows isolating the effect of coordinating capacity requirements from that of cross-border participation. The resulting capacity demand forms the basis for all scenarios considering cross-border participation.}
\item \textit{CM-Implicit}: A capacity market with implicit cross-border participation. \textcolor{R1}{\sout{To determine capacity demand, expected scarcity imports from neighbouring zones are subtracted from the peak residual demand in each zone. This creates a circular dependency, as imports depend on available capacity. We use the investment outcome of the \textit{CM-FBMC} to determine available imports, thereby enabling comparability with other scenarios}Each zone's capacity demand is that determined in the joint scarcity assessment minus the average imports it receives in the scarcity scenarios under scenario \textit{CM-ref}. }
\item \textit{CM-NTC}: A capacity market with explicit cross-border participation limited by a pre-defined maximum entry capacity (MEC) on each border. \textcolor{R3}{\sout{}We use capacity demands from scenario \textit{CM-ref}.} For comparability, we \textcolor{R3}{\sout{set these}determine best-case} MECs using the investment decisions from the \textit{CM-FBMC} scenario, following the procedure outlined in Section \ref{sec:cap-NTC}. \textcolor{R3}{\sout{}Thus, results from this scenario should be viewed as a best-case NTC, since we exclude some of the investment distortions that would arise from an endogenously determined MEC.} As a consequence, results \textcolor{R3}{\sout{reflect only the difference}differ from \textit{CM-FBMC} only} in the treatment of cross-border capacity trade.
\item \textit{CM-FBMC}: \textcolor{R1}{\sout{A}The proposed} flow-based coupling of \textcolor{R1}{\sout{the capacity market}capacity markets}. \textcolor{R1}{We use capacity demands obtained in scenarios \textit{CM-ref}.} The participation of foreign capacity in each zone's capacity market is constrained by the permissible flows across the system under predefined scarcity scenarios \textcolor{R1}{\sout{}(Figure \ref{fig:flow_based_capacity_market}). \textcolor{R1}{\sout{We use capacity demands obtained in scenarios \textit{CM-ref}}. Scarcity events depict the time steps in which ENS occurs under \textit{EOM-cap}. We represent projected scarcity events as (i) "simultaneous scarcity", where all zones demand 95~percent of their peak residual demand simultaneously, and (ii) "zonal scarcity", where the zone experiencing critical scarcity demands 100~per~cent of its peak residual demand, while neighbouring zones each simultaneously demand 90~per~cent of their peak residual demand.} The scarcity scenarios are identical to those used in the joint scarcity assessment.}
\end{enumerate}

\begin{table}[!t]
\caption{Summary of the market designs analysed, indicating whether a capacity market (CM) is implemented, the type of cross-border participation (CBP), and the willingness-to-pay (WTP) or price-cap (\euro/MWh) in the energy market.}
\label{tab:scenario}
\centering
\scriptsize
\setlength{\tabcolsep}{8pt}
\begin{tabular}{lcccl}
\toprule
Scenario & CM & CBP & WTP / Price-cap\\
\midrule
EOM-ref      & \xmark  & \xmark                   & 20,000 \\
EOM-cap      & \xmark  & \xmark                   & 4,000  \\
CM-NoCBP     & \cmark  & \xmark                   & 4,000  \\
CM-FBMC      & \cmark  & Explicit - Flow-based     & 4,000  \\
CM-NTC       & \cmark  & Explicit - MEC           & 4,000  \\
CM-Implicit  & \cmark  & Implicit                 & 4,000  \\
\bottomrule
\end{tabular}
\scriptsize
\end{table}

\begin{table}[!ht]
    \centering
    \caption{Residual peak demand valued at the price cap (MW) in each zone during observed scarcity scenarios and zonal capacity demands (MW) per capacity market design.}
    \label{tab:scarcity_scenarios}
    \begin{subtable}[b]{0.5\textwidth}
    \scriptsize
    \centering
    \caption{Residual peak demand valued at the price cap (MW)} 
    \setlength{\tabcolsep}{9pt}
    \begin{tabular}{lrrrr}
        \toprule
        Scenario & Zone A & Zone B & Zone C & Total\\
        \midrule
        Scarcity at A & \textbf{16,870} & 13,293 & 14,722 & 44,885 \\
        Scarcity at B & 16,006 & \textbf{14,829} & 14,770 & \textbf{45,605} \\
        Scarcity at C & 15,046 & 13,293 & \textbf{16,450} & 44,789 \\
        \bottomrule
    \end{tabular}
    \end{subtable}
\\
\vspace{\baselineskip} 
\begin{subtable}[b]{0.5\textwidth}
\centering
\setlength{\tabcolsep}{5pt}
\scriptsize
\caption{Capacity demand (MW)} 
\begin{tabular}{@{}lrrrr@{}}
\toprule
\text{Scenario} & \text{Zone A} & \text{Zone B} & \text{Zone C} & \text{Total} \\
\midrule
CM-NoCBP                & 16,870 & 14,829 & 16,450 & 48,149 \\
CM-ref, CM-FBMC, CM-NTC & 16,156 & 13,957 & 15,493 & 45,606 \\
CM-Implicit             & 14,778 & 13,084 & 14,238 & 42,100 \\
\bottomrule
\end{tabular}
\end{subtable}
\end{table}

\textcolor{R3}{
All code and input data are available online \cite{Adekola2026CodeSupplement}.}

Below, we present the resulting capacity mixes (Section~\ref{sec:capacity}). Next, we discuss the implications for cross-border capacity trade and congestion rent (Section~\ref{sec:interconnectors}). Lastly, we reflect on the system costs and the costs to consumers (Section~\ref{sec:consumers}).

\subsection{Investment decisions and capacity mix}\label{sec:capacity}
\begin{figure}[!t]
\centering
\begin{subfigure}[b]{1.0\linewidth}
\includegraphics[width=1.0\linewidth]{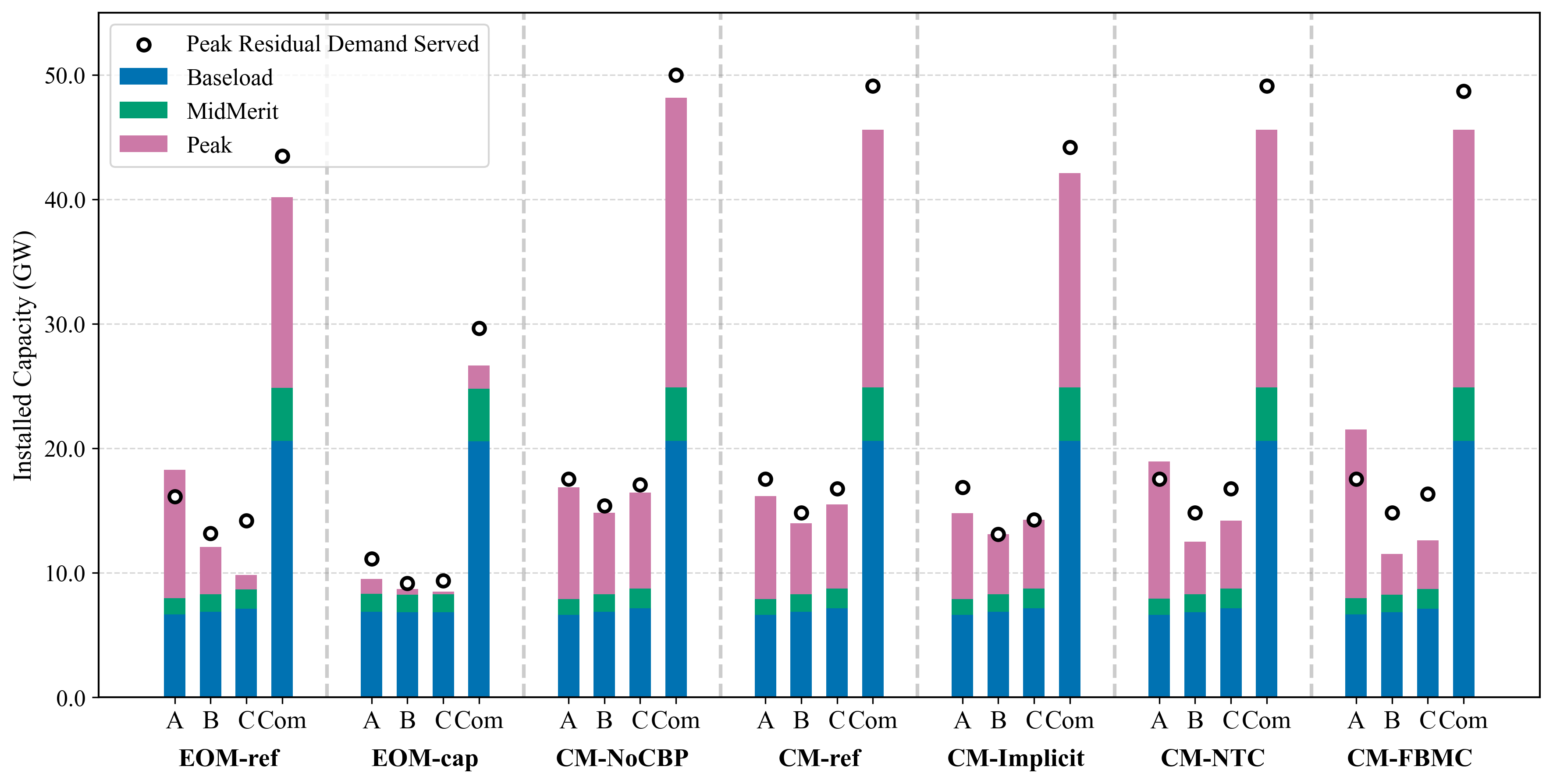}
\caption{Capacity mix by zone and market design. "Com" indicates the aggregated result across all zones. } 
\label{fig:capacity_tot}
\end{subfigure} 
\begin{subfigure}[b]{1.0\linewidth}
\includegraphics[width=1.0\linewidth]{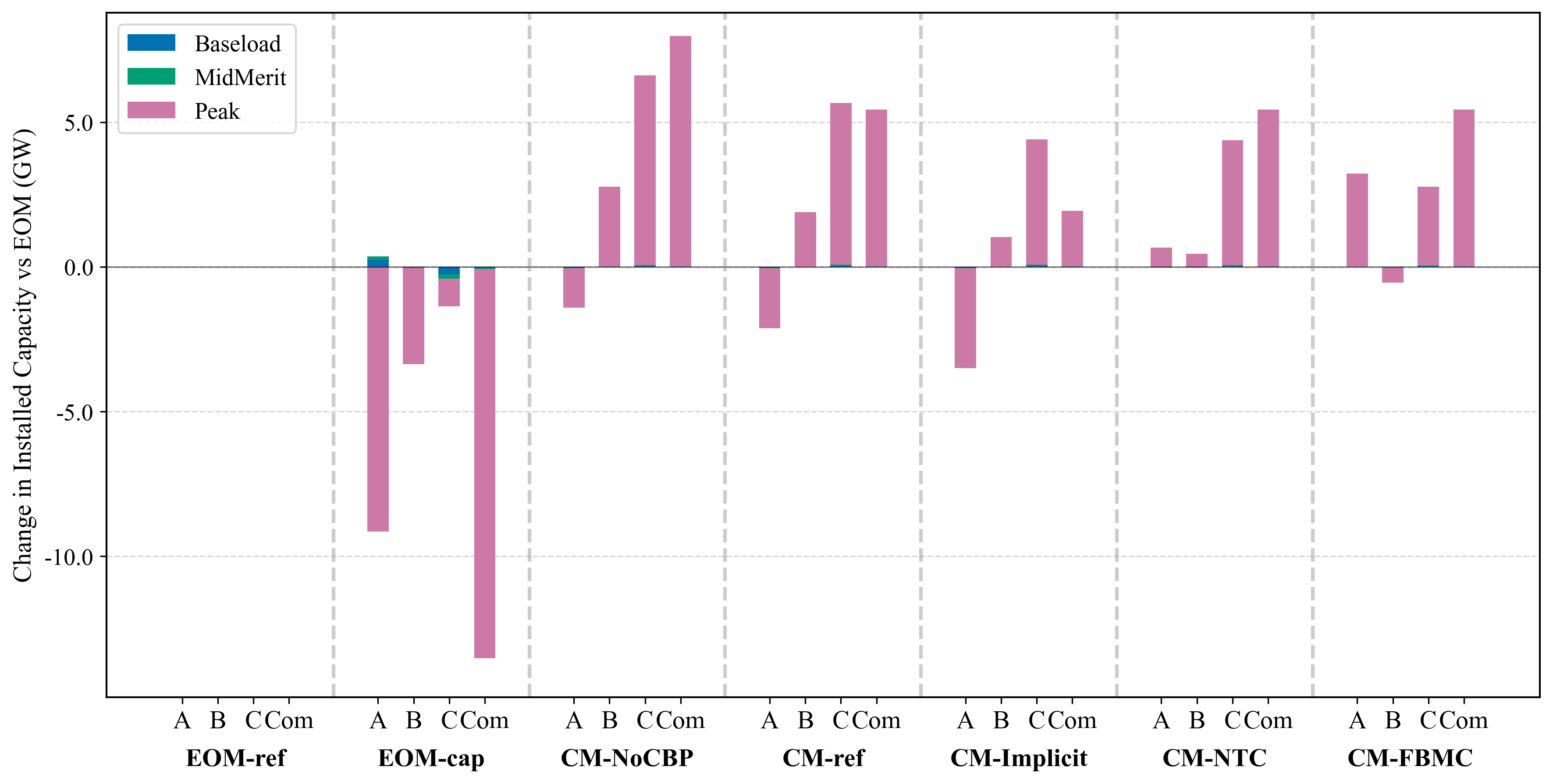}
\caption{Change in installed capacity w.r.t. \textit{EOM-ref}.}
\label{fig:capacity_diff}
\end{subfigure}
\caption{Capacity markets restore investment levels in peaking units but risk over-procurement without coordinated cross-border participation; \textcolor{R1}{\sout{our proposed flow-based coupling (\textit{CM-FBMC}) limits total build compared to \textit{CM-NTC}. CM-NoCBP, CM-FBMC and CM-NTC reallocate investments across zones compared to CM-implicit}\textit{EOM-cap} and capacity markets without explicit cross-border capacity remuneration (\textit{CM-NoCBP}, \textit{CM-ref}, \textit{CM-implicit}) displace capacity away from cheaper Zone~A relative to \textit{EOM-ref}, while the explicit designs relocate capacity to Zone~A}.}
\label{fig:capacity}
\end{figure}

In the reference Energy-Only Market (\textit{EOM-ref}), investment is driven by scarcity pricing. 
\textcolor{R1}{In this idealised \textit{EOM-ref}, prices may rise to the WTP of 20{,}000~\euro/MWh, at which the entire price-responsive share of demand ($20\%$ of the reference level) would voluntarily curtail. However, in equilibrium scarcity prices reach at most 16{,}608~\euro/MWh. Realised peak demand falls to $17.5$, $13.7$ and $15~\mathrm{GW}$ in zones A, B and C, from reference peaks of $19$, $16$ and $18~\mathrm{GW}$.}
Implementing a price cap (\textit{EOM-cap}) \textcolor{R1}{\sout{limits these scarcity rents} truncates prices at 4{,}000~\euro/MWh, with consequences on both consumer demand and generation capacity investments. The forgone scarcity rents}\textcolor{R1}{\sout{This results in} cause} a decline in investment, primarily in peaking units (Fig. \ref{fig:capacity}). Total capacity falls by 13.5~GW relative to the reference \textcolor{R1}{\sout{}energy-only market}. \textcolor{R1}{\sout{}On the consumer side, the volume of elastic demand that can express its willingness to pay in the market decreases from 20\% to 4\% of the reference level. The combination of less firm capacity and less price-responsive demand visible in the market results in} Energy Not Served ($\mathrm{ENS}$)\textcolor{R1}{\sout{reaches}}
\textcolor{R1}{\sout{$41.2~\mathrm{GWh}$, $30~\mathrm{GWh}$, and $33~\mathrm{GWh}$}of $38.9~\mathrm{GWh}$, $28.9~\mathrm{GWh}$, and $37.3~\mathrm{GWh}$ in zones A, B, and C.}

A capacity market without cross-border participation (\textit{CM-NoCBP}) successfully eliminates $\mathrm{ENS}$; however, it does so through autarky. In line with observations in the literature \cite{Mays2019AsymmetricMarkets}, the system sees a surge in low-capex, high-opex peaking units (pink bars), increasing total capacity by \textcolor{R1}{\sout{10~GW}8~GW} w.r.t. \textit{EOM-ref}\textcolor{R1}{, indicating overprocurement}. We observe a capacity displacement of  \textcolor{R1}{\sout{600~MW}1.4~GW} away from Zone A w.r.t \textit{EOM-ref}, as the isolation prevents Zone A from capitalising on its export potential. To mitigate this over-procurement \textcolor{R1}{and capacity displacement}, we \textcolor{R1}{\sout{}carry out a joint scarcity assessment in \textit{CM-ref}, and further} introduce cross-border participation under three designs: \textit{CM-implicit}, \textit{CM-NTC}, and \textit{CM-FBMC}.

\textcolor{R4}{Coordinating capacity requirements through the joint scarcity assessment eliminates unserved energy with less installed capacity than in \textit{CM-NoCBP}. Zone~B, for example, procures 872~MW less than its own peak residual demand and covers the gap with 874~MW of imports at its peak. The flows presumed via the joint assessment, therefore, materialise in the energy market, though \textit{CM-ref} neither remunerates nor secures them. Coordination, however, does not relocate investment. Without cross-border participation, the capacity displaced from Zone~A w.r.t.\ \textit{EOM-ref} grows to 2.1~GW, above the 1.4~GW under \textit{CM-NoCBP}, while Zones B and C add 1.9 and 5.7~GW (Fig.~\ref{fig:capacity}). In total, \textit{CM-ref} installs 2.5~GW below \textit{CM-NoCBP}, with the reduction spread across all zones (0.7, 0.9 and 0.9~GW in Zones A, B and C).}

\textcolor{R4}{\sout{By deducting expected foreign imports from demand without remuneration, the implicit cross-border participation \textit{CM-implicit} minimises investment, adding only 4~GW w.r.t. \textit{EOM-ref} (a 6~GW reduction compared to the \textit{CM-NoCBP}), with some unserved energy re-emerging (4.2~GWh across the system). The latter is to be expected, as capacity demands were set assuming the investment levels in neighbouring zones from the \textit{CM-FBMC} scenario. However, without explicit remuneration, these investments do not materialise, resulting in the unavailability of the expected imports. A capacity displacement of 600~MW from Zone A w.r.t \textit{EOM-ref} is observed, similar to \textit{CM-NoCBP} scenario.}
Implicit cross-border participation \textit{CM-implicit} deducts the 3.5~GW of imports that arise in each zone's scarcity period under \textit{CM-ref} (1.4, 0.9 and 1.3~GW in Zones A, B and C) without remunerating them. However, the unremunerated import capacity does not materialise because, without explicit remuneration, the foreign capacity expected to support these imports is not built. Total installed capacity falls to 42.1~GW, 1.9~GW above \textit{EOM-ref} and 6.0~GW below \textit{CM-NoCBP}. Consequently, unserved energy re-emerges across all zones (12.4~GWh system-wide). Zone~A bears the largest capacity displacement w.r.t \textit{EOM-ref} with 3.5~GW less capacity investment, as capacity it otherwise would build for export is left unremunerated.}

Explicit cross-border participation under both \textit{CM-NTC} and \textit{CM-FBMC} \textcolor{R1}{\sout{utilises coordinated scarcity assessment, thus reducing system-wide over-procurement w.r.t. \textit{CM-NoCBP} by 2.5~GW while serving the same peak residual demand}
starts from the same joint scarcity assessment as \textit{CM-ref}, thus total installed capacity stays at 45.6~GW, 2.5~GW below \textit{CM-NoCBP}, and unserved energy remains negligible, at less than $10^{-4}$\,\% of served annual demand in the system}.
\textcolor{R1}{\sout{Unlike \textit{CM-implicit} and \textit{CM-NoCBP}, no displacement of capacity away from Zone~A w.r.t. \textit{EOM-ref} is observed. Both \textit{CM-NTC} and \textit{CM-FBMC} shift capacity investment toward lower-cost and system-friendly locations to varying degrees. When compared to \textit{CM-NTC}, \textit{CM-FBMC} better optimises location, shifting 1.3~GW of investment away from the costlier Zone C to the lower-cost Zone A.}
As foreign capacity obligations are remunerated in the explicit designs, Zone~A with lower cost of new entry exploits this opportunity and builds 2.8~GW (\textit{CM-NTC}) and 5.3~GW (\textit{CM-FBMC}) above its own capacity demand to serve capacity obligations remunerated by Zones~B and C. This is also observed in the obligations it exports (Fig.~\ref{fig:net_capacity_trade}). Comparing both explicit designs, \textit{CM-FBMC} relocates more capacity, shifting 2.6~GW more of investment into the lower-cost Zone~A relative to \textit{CM-NTC}.} \textcolor{R4}{\sout{}Recall that we base the maximum entry capacities used in \textit{CM-NTC} on the available capacity in the \textit{CM-FBMC}. As such, the result excludes the investment distortions that would result from using the NTC approach and can be interpreted as an upper bound for \textit{CM-NTC}.}

Notably, all capacity market designs result in a total capacity that exceeds the \textit{EOM-ref}. While \textit{EOM-cap} under-invests by $13.5$~GW, capacity mechanisms add \textcolor{R1}{\sout{$4\text{--}10$~GW}$1.9\text{--}8$~GW} above \textit{EOM-ref} level. This arises from feedback between electricity prices and the elastic electricity demand. In \textit{EOM-ref}, high scarcity prices reduce electricity consumption \textcolor{R1}{\sout{}through self-rationing by consumers whose WTP is below the market-clearing price}, and consequently, the peak residual demand served is reduced to 44~GW. \textcolor{R1}{\sout{Capacity markets suppress these price spikes (Section \ref{sec:interconnectors}) and the associated demand elasticity, with the peak residual demand served increasing to 48--50~GW}The price cap truncates electricity prices (Section \ref{sec:interconnectors}), removing the efficient self-rationing of consumers with demand valued above the price cap (4{,}000~\euro/MWh).
\sout{Consequently,} To avoid involuntary curtailment of demand, more firm capacity is required\sout{ to guarantee the same level of reliability}}. \textcolor{R1}{As a result, the peak residual demand served rises from 44~GW in \textit{EOM-ref} to  49--50~GW in the scenarios considering a capacity market}.

\subsection{Cross-border trades and congestion revenues} \label{sec:interconnectors}
\begin{figure}[!t]
\centering
\includegraphics[width=1.0\linewidth]{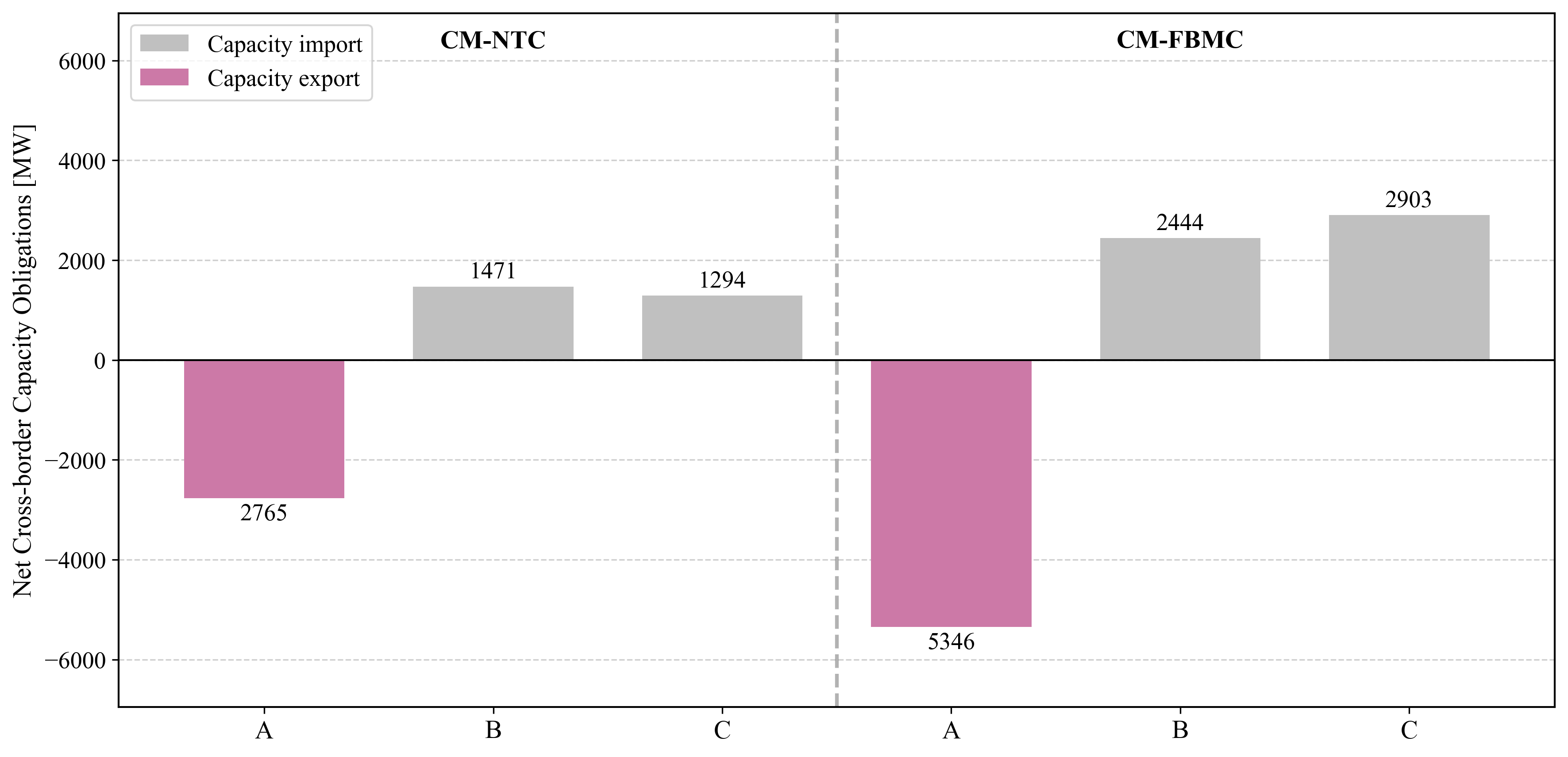}
\caption{Net cross-border capacity trade by zone.}
\label{fig:net_capacity_trade}
\end{figure}

\textcolor{R1}{\sout{The total cross-border capacity trade is 1.4~GW lower under \textit{CM-NTC} than under \textit{CM-FBMC} (Figure \ref{fig:net_capacity_trade}). In the NTC design, Zone~A is the sole net exporter and provides 3.5~GW of capacity obligations to its neighbours. Zones~B and~C are net importers, 369~MW and 3~GW. Under \textit{CM-FBMC}, Zone~A exports 4.8~GW. This allows Zone~C to reduce local investment. Zone~C instead imports 4.4~GW of obligations from Zone~A—an increase of 1.3~GW over the NTC case. Zone~B remains a net importer with 408~MW. Higher trade under \textit{CM-FBMC} does not compromise physical deliverability. For example, Zone~C successfully imports 5~GWh during scarcity}
Cross-border trade of capacity obligations is 2.6~GW higher under \textit{CM-FBMC} than under \textit{CM-NTC} (Figure~\ref{fig:net_capacity_trade}).} \textcolor{R3}{In \textit{CM-NTC},  bilateral transfer limits determined based on investment levels that materialise in \textit{CM-FBMC} are $A \leftrightarrow B\colon [740.3, -620]$\text{ MW}, $A \leftrightarrow C\colon [574.3, -454]$\text{ MW}, and $B \leftrightarrow C\colon [694, -694]$\text{ MW}.} \textcolor{R1}{Zone~A is the sole net exporter in both designs. Under \textit{CM-NTC} it sells 2.8~GW of obligations, of which Zones~B and~C procure 1.5 and 1.3~GW. Under \textit{CM-FBMC} it sells 5.3~GW, and both neighbours roughly double their foreign procurement, to 2.4~GW (Zone~B) and 2.9~GW (Zone~C), reducing within-zone investment by the same amounts. The higher cross-border capacity trade is shown to be physically deliverable during scarcity with Zone~C importing 3.7~GWh/h, against 2.5~GWh/h under \textit{CM-NTC}}. 

We show the network operator's congestion rents in the energy and capacity market in Table~\ref{tab:congestion_rent}. In the reference Energy-only market (\textit{EOM-ref}), high price differentials during scarcity drive congestion rents of \textcolor{R1}{\sout{72.4~M\euro}73.1~M\euro}. During congestion, prices peak at \textcolor{R1}{\sout{14,057 \euro/MWh}14,100 \euro/MWh} in Zone~A and 16,608 \euro/MWh in Zone~C. These spreads generate high revenue. Implementing a price cap (\textit{EOM-cap}) drastically reduces these rents to \textcolor{R1}{\sout{1.1~M\euro}2.6~M\euro}. The cap artificially reduces flows (due to under-investment across all zones) and the price spread between zones, as all zones hit the 4,000 \euro/MWh cap simultaneously. \textcolor{R3}{\sout{\textit{CM-Implicit} follows the same trend; revenue there falls to 0.1~M\euro. Finally, in \textit{CM-NoCBP}, over-procurement of capacity leads to lower flows and high price convergence across zones. Congestion revenues collapse to 0.5~M\euro}
The capacity market designs without explicit cross-border participation further erase these rents, albeit for different reasons. In \textit{CM-NoCBP}, overprocurement suppresses scarcity pricing. With prices peaking at 560--650~\euro/MWh in all zones, spreads are minimal and congestion rents collapse to 0.5~M\euro. In \textit{CM-ref} and \textit{CM-Implicit}, scarcity prices reach the cap; however, similar to \textit{EOM-cap}, prices are near the cap in all zones at the same periods, lowering price spreads and congestion rents fall to 0.1 and 0.5~M\euro
}.

Explicitly remunerating cross-border participation restores the value of transmission through "capacity congestion rent". \textcolor{R1}{\sout{ \textit{CM-NTC} generates 64.6~M\euro, partially restoring the levels seen in \textit{EOM-ref}. \textit{CM-FBMC} yields congestion rents of 87.1~M\euro, surpassing the reference market level}
\textit{CM-NTC} generates 39.6~M\euro, most of it in the capacity market, restoring more than half of the \textit{EOM-ref} level. \textit{CM-FBMC} yields 78.7~M\euro, narrowly surpassing \textit{EOM-ref}. Unlike \textit{CM-NTC}, it earns a third of this in the energy market, as greater scarcity exports from Zone~A congest interconnectors, causing price spreads to persist during scarcity.
}

\begin{table}[!t]
\caption{The network operator's congestion rents (M\euro/yr) across market designs.}
\label{tab:congestion_rent}
\centering
\scriptsize
\setlength{\tabcolsep}{1pt}
\begin{tabular}{@{}lccccccc@{}}
\toprule
\text{Market} & \text{EOM-ref} & \text{EOM-cap} & \text{CM-NoCBP} & \text{CM-ref} & \text{CM-Implicit} & \text{CM-NTC} & \text{CM-FBMC} \\
\midrule
Energy   & 73.07 & 2.64 & 0.50 & 0.07 & 0.53 & 0.47  & 32.22 \\
Capacity & 0.00  & 0.00 & 0.00 & 0.00 & 0.00 & 39.08 & 46.46 \\
\midrule
Total revenues & 73.07 & 2.64 & 0.50 & 0.07 & 0.53 & 39.55 & 78.68 \\
\bottomrule
\end{tabular}
\end{table}
%
\subsection{System costs\textcolor{R2}{\sout{}, social welfare} and costs to consumers}\label{sec:consumers}
\subsubsection{System Costs}\label{sec:system_costs}
By design, the energy-only market \textit{EOM-ref} presents the least-cost market design, with a total system cost of 15,398~M\euro{}/yr (Table~\ref{tab:system_costs}).\footnote{Total system costs equal the sum of generation costs, investment costs and cost of lost load. The cost of lost load is the volume of ENS valued at 18{,}667~\euro/MWh; the average value of the demand that is randomly curtailed at the price cap of 4{,}000~\euro/MWh.}
The introduction of a price cap (\textit{EOM-cap}) suppresses investment costs by 870~M\euro{}/yr due to the "missing money" problem, resulting in a surge in the cost of unserved energy \textcolor{R1}{\sout{(2,103~M\euro)}(1{,}963~M\euro/yr)}. Consequently, the \textit{EOM-cap} is the least efficient design, incurring a cost increase of \textcolor{R1}{\sout{nearly 8}7}$\%$ relative to the reference case.

Capacity markets \textcolor{R1}{\sout{eliminate}reduce} ENS costs but increase investment costs. When compared to \textit{EOM-ref}, \textit{CM-NoCBP} increases investment costs by \textcolor{R1}{\sout{over 780~M\euro, or 5.2$\%$}640~M\euro/yr, increasing total system cost by 4.2$\%$}. 
\textcolor{R4}{\sout{The \textit{CM-NTC} design incurs a lower investment cost than the isolated case, but still 3.62$\%$ above the reference case. \textit{CM-FBMC} brings the total costs down to 15,929~M\euro{}. This reduces the cost increase over \textit{EOM-ref} to 3.45$\%$. The \textit{CM-Implicit} achieves the lowest cost among the capacity mechanisms (a 2.6$\%$ increase in cost). Recall, however, that ENS emerges, unlike the explicit designs, which guarantee adequacy. The \textit{CM-Implicit} costs are thus highly sensitive to the cost of unserved energy.}
Coordinating the capacity requirement in CM-ref reduces the required capacity investments by 2.5~GW and lowers annualised investment costs by 181~M\euro{}/yr, limiting the total system-cost increase relative to \textit{EOM-ref} to 3.1\% (470~M\euro{}/yr). \textit{CM-Implicit} reduces investment costs by a further 245~M\euro{}/yr, but this is largely offset by 232~M\euro{}/yr in unserved-energy costs, leaving total system costs only marginally below \textit{CM-ref} and 2.9\% above \textit{EOM-ref}. Unserved energy under \textit{CM-Implicit} is 12.4~GWh per year, equivalent to roughly 0.005\% of annual demand (Table~\ref{tab:system_costs}). \textit{CM-implicit} is sensitive to assumptions on the availability of imports during scarcity. Higher import assumptions would increase the volume of unserved energy, and because ENS costs rise faster than investment savings, the total system cost would increase. Without import deductions, the design collapses to \textit{CM-ref}.}

\textcolor{R3}{Explicit cross-border participation lowers total costs by relocating peaking capacity from zones with a high investment cost to lower-cost zones while keeping ENS below 0.25 GWh/yr. \textit{CM-NTC} reduces investment costs by 40~M\euro{}/yr relative to \textit{CM-ref} (0.7\,\% of 5\,820~M\euro{}/yr), and by
82~M\euro{}/yr (1.4\,\%) under \textit{CM-FBMC}. It removes the unserved energy of
\textit{CM-Implicit}, with ENS falling to 253~MWh/yr (below
$10^{-4}$\,\% of served annual demand) in \textit{CM-FBMC} and to 0.5~MWh/yr in \textit{CM-NTC}. \textit{CM-FBMC} is the least-cost capacity-market design, 258~M\euro{}/yr below \textit{CM-NoCBP}, 55~M\euro{}/yr below \textit{CM-Implicit}, and 35~M\euro{}/yr (or 0.22\%) below a best-case NTC (\textit{CM-NTC}), at a total annualised cost of 15,791~M\euro/yr.}

\begin{table}[!t]
\centering
\caption{\textcolor{R3}{Served demand (GWh/yr), unserved energy (ENS -- GWh/yr), system costs (M\euro/yr) and
cost increase w.r.t.\ the reference EOM case (\textit{EOM-ref}).}}
\label{tab:system_costs}
\scriptsize
\setlength{\tabcolsep}{5pt}
\begin{tabular}{@{}lccccccc@{}}
\toprule
Market & Demand & ENS & Gen.\ & Inv.\ & ENS & Total & Increase \\  
Design& Served & Volume & Cost &  Cost & Cost & Cost & (\%) \\
\midrule
EOM-ref     & 271,630 & 0        & 10,036 & 5,361 & 0     & 15,397 & --   \\
EOM-cap     & 271,532 & 105.14  & 10,015 & 4,491 & 1,963 & 16,468 & 6.96 \\
\midrule
CM-NoCBP    & 271,664 & 0        & 10,048 & 6,001 & 0     & 16,049 & 4.23 \\
CM-ref      & 271,659 & 0.01    & 10,046 & 5,820 & 0.1     & 15,866 & 3.05 \\
CM-Implicit & 271,641 & 12.42   & 10,038 & 5,575 & 232   & 15,846 & 2.91 \\
CM-NTC      & 271,659 & $<$0.01 & 10,046 & 5,780 & 0     & 15,826 & 2.78 \\
CM-FBMC     & 271,659 & 0.25    & 10,048 & 5,738 & 4.7     & 15,791 & 2.56 \\
\bottomrule
\end{tabular}
\end{table}
\subsubsection{\textcolor{R2}{Social welfare}}\label{sec:social_welfare}
\textcolor{R2}{We decompose the change in welfare w.r.t \textit{EOM-ref} in changes in consumer surplus, producer surplus and congestion rents (Figure~\ref{fig:welfare_decomposition}).\footnote{\textcolor{R2}{To compute the change in consumer surplus, we re-evaluate energy market outcomes using the uncapped demand curve. The obtained consumer surplus is corrected with the cost of energy not served and capacity payments.}} \textit{EOM-ref} leaves market-clearing price uncapped, enabling efficient self-rationing by consumers with a WTP below the clearing price, and providing a reference value for surplus generated in the market. The largest welfare loss is observed in  \textit{EOM-cap} ($-864$~M\euro/yr), the lowest in \textit{CM-FBMC} ($-171$~M\euro/yr), consistent with what is obtained in (Section~\ref{sec:system_costs}). Under the price cap, consumers bear the loss ($-982$~M\euro/yr, driven by the unserved energy), while producers gain 189~M\euro/yr as inframarginal rents rise, particularly for renewables and baseload generators. Capacity markets reverse this trend. Producer surplus decreases as capacity revenues compensate investment cost but lower inframarginal rents. However, it significantly improves consumer surplus. Only \textit{CM-NoCBP} ($-125$~M\euro/yr, due to over-procurement) and \textit{CM-Implicit} ($-17$~M\euro/yr, energy-not-served) leave consumers worse off.}

\textcolor{R2}{We distinguish two sources of welfare gains. First, coordinating the capacity demand through a joint scarcity assessment (\textit{CM-NoCBP} to \textit{CM-ref}) improves welfare by 172~M\euro/yr, with most accruing to consumers. Secondly, explicit cross-border participation adds 41~M\euro/yr under \textit{CM-NTC} and 75~M\euro/yr under \textit{CM-FBMC}, which materialises almost entirely as congestion revenue. \textit{CM-FBMC} is the only design that fully restores the value of interconnection compared to \textit{EOM-ref}.
}

\subsubsection{Cost to consumers}\label{sec:consumer_costs}
\begin{figure}[!t]
\centering
\begin{subfigure}[b]{1.0\linewidth}
\includegraphics[width=1.0\linewidth]{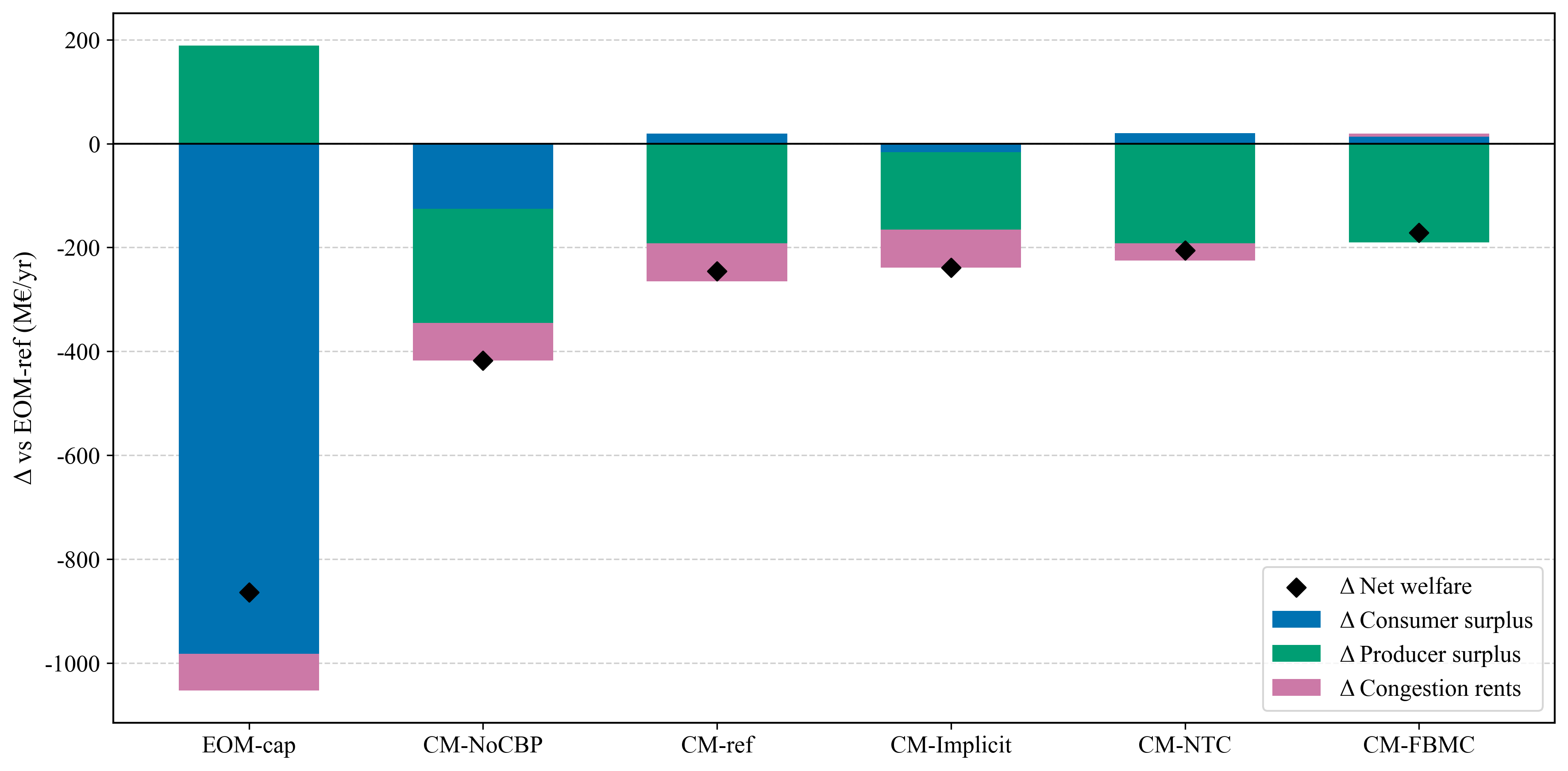}
\caption{\textcolor{R2}{Welfare decomposition showing the change w.r.t \textit{EOM-ref} in consumer and producer surplus, congestion rents and the net welfare across market designs}}
\label{fig:welfare_decomposition}
\end{subfigure}
\begin{subfigure}[b]{1.0\linewidth}
\includegraphics[width=1.0\linewidth]{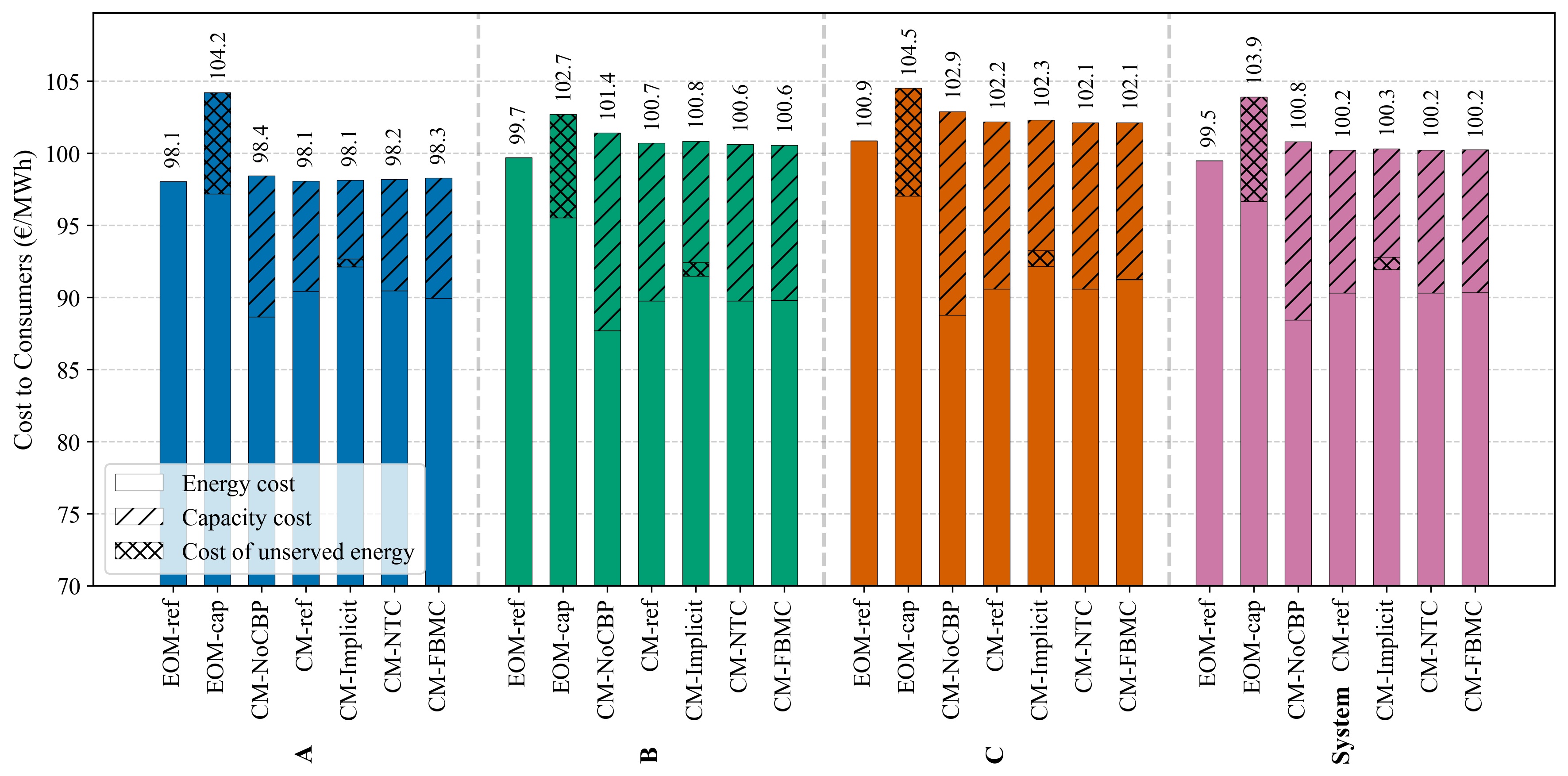}
\caption{Average consumer cost by zone, decomposed into energy, capacity, and ENS (\euro/MWh). “System” represents the average cost of all zones and enables comparison of total system costs.}
\label{fig:average_cost}
\end{subfigure}
\caption{\textcolor{R1}{\sout{}We show the impact of cross‑border capacity participation and coupling design on (a) Welfare decomposition [M\euro/yr] and (b) total average cost [\euro/MWh] in Zones A–C.}} \label{fig:cost}
\end{figure}

Figure~\ref{fig:average_cost} summarises the cost of electricity for consumers normalised by the demand served (\euro/MWh), decomposed into three components: the energy cost, cost of energy not served and capacity cost.

In \textit{EOM-ref}, average costs reach $99.7$~\euro/MWh in Zone~B and $100.9$~\euro/MWh in Zone~C. Zone~A sees a lower cost of $98.1$~\euro/MWh. The system-wide average is $99.5$~\euro/MWh. Neither a capacity market nor unserved energy exists in this reference case. Introducing a price cap increases consumer costs across all zones because unserved energy emerges. System-wide costs rise from $99.5$~\euro/MWh in \textit{EOM-ref} to \textcolor{R1}{\sout{$104.4$~\euro/MWh}$103.9$~\euro/MWh} in \textit{EOM-cap}. The cap limits scarcity prices but triggers underinvestment and ENS costs. These effects are shown in the hatched areas of Figure~\ref{fig:average_cost}.

Capacity markets provide reliability at a premium. \textit{CM-NoCBP} eliminates ENS; however, capacity payments represent \textcolor{R1}{\sout{$12.8$\% of the total cost. Capacity market prices in Zones A, B, and C clear at the costs of new entry: $60{,}000$~\euro/MW, $70{,}000$~\euro/MW, and $80{,}000$~\euro/MW. The system-level average cost falls to $101.3$~\euro/MWh, below the \textit{EOM-cap} but still above the reference Energy-only market}$12.3$\% of the total cost to consumers. Capacity market prices in Zones A, B, and C clear at or below the costs of a new build: $59{,}777$~\euro/MW/yr, $69{,}623$~\euro/MW/yr, and $80{,}000$~\euro/MW/yr. The system-level average cost to consumers falls to $100.8$~\euro/MWh, $3.1$~\euro/MWh below the \textit{EOM-cap} and $1.3$~\euro/MWh above the reference Energy-only market.
Coordinating capacity demands through a joint scarcity assessment in \textit{CM-ref} lowers capacity prices by roughly 10,000~\euro/MW/yr in every zone (to 48,904, 58,965 and 69,886~\euro/MW/yr) and reduces capacity payments to 9.9\% of the cost. Consumers see a net saving of 0.6~\euro/MWh}.

\textcolor{R1}{\sout{In \textit{CM-Implicit}, zones account for foreign support without paying for it explicitly, thus reducing system-wide average costs to $100.0$~\euro/MWh. Zone~C sees its average cost drop to  $99.9$~\euro/MWh, $1$~\euro/MWh below \textit{EOM-ref}. Capacity costs fall to 8.9\% of the total, with capacity clearing prices ranging from $45{,}956$~\euro/MW to $65{,}323$~\euro/MW, reflecting lower demand.}
\textit{CM-Implicit} goes further. Zones count on imports from foreign capacity without remunerating them. Capacity payments fall to 7.5\% of the cost (38,254--59,246~\euro/MW/yr), but the imports do not materialise during scarcity events. Unserved energy adds 0.9~\euro/MWh and higher energy costs a further 1.6~\euro/MWh, leaving consumers worse off than under \textit{CM-ref}.}

Under \textit{CM-FBMC}, consumers face the same average system cost as in \textit{CM-NTC}. \textcolor{R1}{\sout{However, the flow-based approach reduces capacity prices in Zones B and C by $1{,}019$~\euro/MW and $2{,}275$~\euro/MW compared to the NTC case. As a result, capacity prices as a percentage of total costs decline from $12.2$\% to $12$\%.}
Explicit cross-border participation (\textit{CM-NTC} and \textit{CM-FBMC}) leaves the system-wide cost at the \textit{CM-ref} level, but redistributes it across zones. Zone~A's consumers face a higher capacity cost due to demand from neighbouring zones while Zones~B and~C pay 0.1~\euro/MWh less. Between both designs, \textit{CM-FBMC} lowers capacity prices in Zones~B and~C by 390 and 4,076~\euro/MW/yr and raises Zone~A's by 3,876~\euro/MW/yr.
}

\section{Discussion}\label{sec:discussion}
We propose a flow-based coupling of European capacity markets (\textit{CM-FBMC}), representing a significant departure from current arrangements. Below, we reflect on its implications and on the challenges it may raise.


One of the most significant hurdles to regional capacity markets is presented by the unique needs of Member States. Political reluctance to relinquish control over national resource adequacy and capacity mix hinders progress toward regional integration of capacity markets. Our proposed approach avoids this conflict. It does so by separating long-term investment needs and short-term adequacy. Member States still determine long-term (multi-year) instruments to de-risk investment in capital-intensive assets. The coupled annual capacity markets then provide an opportunity to re-trade capacity obligations held by governments and facilitate investment in fast-to-deploy assets. Each MS may define its own long-term contract structure and parameters of the capacity demand curve (e.g. the demand curve may be defined "centrally" or "decentrally") -- only the participation in annual capacity auctions, the traded contract (e.g., reliability options), and coincidental scarcity assessments need to be harmonised.

Even as we accommodate national preferences, the proposed mechanism continues to guide efficient regional investment. It reduces overprocurement by adopting a regional adequacy assessment that exploits the degree of non-simultaneity in scarcity events. A common view of scarcity scenarios across the system incentivises investment levels only up to the peak simultaneous scarcity across zones, rather than the sum of individual zonal peaks. As we demonstrate in our case study \ref{sec:capacity}, efficiency gains in \textit{CM-FBMC} are driven both by the mechanism's ability to exploit the degree of non-simultaneity in scarcity events within the network and by relocating investment to cheaper locations. 
\textcolor{R2}{\sout{Two factors may diminish these efficiency gains. Ill-defined scarcity expectations can lead to inaccurate capacity procurement. Furthermore, if MS ignore price signals and promotes investment in the wrong locations (i.e., non-deliverable locations), trade in the annual, coupled capacity market remains limited.}}

\textcolor{R2}{\sout{} Two factors may erode these efficiency gains. First, overly conservative regional scarcity expectations at Y-1 may drive inaccurate capacity procurement. Second, several frictions may impede the efficient transmission of price signals between annual, coupled capacity markets and the long-term, national instruments. For example, member states may strategically overprocure firm capacity to guarantee national autonomy or underprocure to free-ride on neighbouring resources.} \textcolor{R2}{The coupled annual auction can disincentivise such behaviour. Member states are mandated to re-market all procured capacity under the national mechanisms into the coupled annual auction. Consequently, the cost of overprocurement is made explicit as the spread between the cost of the long-term contracts and the annual clearing price. Consider a Member State "A" that over-procures multi-year, long-term capacity contracts. Neighbouring zones then benefit from purchasing reliability obligations from "A". If transmission constraints in the coupled capacity auction become binding, capacity prices in zone "A" drop below the cost of some of the long-term contracts held by Member State "A". This makes the cost of overprocurement explicit. Member states may, however, choose to absorb the cost, for example, as the price for autonomy, and if all member states systematically overprocure in the first layer, the outcome becomes similar to that of the \textit{CM-NoCBP}.}

\textcolor{R}{\sout{Nonetheless}Compared to current approaches to cross-border participation in capacity markets}, our design resolves the "circular dependency" inherent in both \textit{CM-NTC} and \textit{CM-Implicit} approaches. Both rely on ex-ante parameterisations -- either in setting Maximum Entry Capacities (\textit{CM-NTC}) or to reduce local capacity targets (\textit{CM-Implicit}). The implicit approach is particularly prone to a "chicken-and-egg" dilemma: it deducts expected imports from local demand without providing remuneration to incentivise investment in the capacity that enables these imports. As such, the assumed surplus may be crowded out in the long run. Our results show this fragility through the re-emergence of unserved energy in the \textit{CM-Implicit} scenario, confirming that reliance on non-remunerated cross-border import may undermine resource adequacy. An increase in the valuation of reliability would sharply penalise the unserved energy risk inherent in the implicit approach, reducing its economic efficiency.

The \textit{CM-FBMC} addresses this issue structurally. 
By using a detailed grid model to clear capacity offers, it ensures that cleared capacity is deliverable w.r.t. network constraints under a set of defined scarcity scenarios. Flow-based domains internalise loop flows, critical network elements, and contingencies, thereby capturing grid physics more accurately than NTC-based “maximum-entry-capacity” caps. FBMC thus requires less conservative assumptions, unleashing more cross-border trade of firm capacity obligations.
Additionally, the flow-based market coupling of annual capacity markets provides explicit price signals for reliability. Zones with adequacy issues will clear their capacity markets at higher prices, providing incentives for cross-border trade and investment in non-capital-intensive assets.

\textcolor{R3}{\sout{Lastly, we discuss limitations of our work, and potential implementation challenges. As TSOs already generate year‑ahead scarcity scenarios and flow-based domains for the European Resource Adequacy Assessment (ERAA), extending these processes to produce harmonised, year-ahead anticipated scarcity scenarios (flow-based domains, availability of generation assets and demand) is the next logical step. Such a role may be assumed by Europe's Regional Coordination Centres (RCC). However, this approach could increase computational effort, as it requires accounting for grid uncertainties and the N-1 security criterion when computing flow-based domains. Furthermore, because flow-based market coupling has been criticised for opacity, its application in capacity markets must prioritise transparency.
Lastly, we discuss \textcolor{R3}{limitations of our work, and} potential implementation challenges.
}
As the proposed design is yet to be implemented in any known real-world setup, we elaborate on the limitations of our work and potential implementation challenges. Our equilibrium model assumes perfect competition and risk-neutral agents in both electricity and capacity markets. Extending the analysis to investigate the effect of risk aversion and market power on cross-border capacity market design constitutes a relevant avenue for future research. Additionally, we certify capacity deliverability under a stylised finite scenario set $\mathcal{S}$. Operationalising the market design would require a robust set of scarcity scenarios. As TSOs already generate year‑ahead scarcity scenarios and flow-based domains for the European Resource Adequacy Assessment (ERAA), extending these processes to produce harmonised, year-ahead anticipated scarcity scenarios (flow-based domains, availability of generation assets and demand) is the next logical step. Such a role may be assumed by Europe's Regional Coordination Centres (RCC).} \textcolor{R3}{Note, however, that our network representation only considers state-$N$ and ignores contingencies during scarcity events. Real-world implementation would account for grid uncertainties through the $N-k$ security criterion. The combination of a scarcity scenario set and $N-k$ contingencies may considerably increase the computational effort required for market clearing. Lastly, because flow-based market coupling has been criticised for opacity, its application in capacity markets must prioritise transparency.}

\section{Conclusion}\label{sec:conclusion}
Capacity markets are becoming a structural element of the European electricity market, but lack harmonisation and cross-border participation necessary for regional efficiency. Our study addresses this issue 
\textcolor{R2}{\sout{through a two-tiered procedure.} by introducing a coordinating layer for national, long-term capacity remuneration mechanisms.} \textcolor{R2}{We propose linking annual, zonal capacity auctions using the flow-based market coupling methodology, in which long-term contracts held by governments can be re-marketed and new assets with short lead times can be attracted. As such,} the coupled annual capacity market \textcolor{R2}{\sout{then allows for compatibility} is compatible} with long-term instruments issued by Member States to mitigate investment risk.

By endogenously modelling network flows in anticipated scarcity scenarios, our approach eliminates the circular dependencies between import assumptions and available capacity in implicit cross-border participation, and between maximum entry capacities and available capacity in current explicit cross-border participation methods, and guarantees the delivery of contracted capacity during those scarcity events.

Our modelling results show that this design more accurately values the contribution of interconnection and \textcolor{R3}{\sout{yields efficiency gains}relocates capacity to cheaper locations under deliverability constraints}. Moreover, our design addresses challenges posed by fragmented national capacity mechanisms while respecting the heterogeneous adequacy requirements of Member States and their sovereign right to steer their capacity mix.

\textcolor{R1}{Future work should focus on detailing the design of the annual auctions and the capacity product to be traded. Key features include the degree of harmonization required across member states and the valuation of resources that have non-symmetric contributions to meeting demand during scarcity situations across zones, such as energy-constrained assets.}
%



\section*{}\label{sec:appendix}
\section*{Appendix I}\label{sec:appendix_1}
\textcolor{R2}{The equilibrium is computed using an exchange-form consensus ADMM implemented in Julia, in which each agent's convex quadratic subproblem is solved in parallel with Gurobi. Each iteration solves Problems~(1)--(4) in parallel, given the prices, then updates $\lamEzt$ and $\lamCmz$ as duals of the zonal imbalances. The algorithm terminates when, for every zone and for both markets (electricity and capacity), the primal and dual residuals fall below an absolute tolerance $\epsilon=0.1$ (in MW for primal residuals; in \euro/MWh and \euro/MW-year for EOM and CM dual residuals). The penalty parameter $\rho$ is adapted per zone and market via residual balancing following Boyd et al~\cite{Boyd2011DistributedMultipliers}. Converged runs typically require 460-2{,}500 iterations (13-38 seconds of wall time), with each iteration taking residuals to decay steadily (Table~\ref{tab:admm_convergence}). All runs are cold-started, with prices, quantities, and penalty terms initialised at zero.}

\textcolor{R2}{Because each agent solves a convex quadratic problem, a converged ADMM solution satisfies the conditions for competitive equilibrium. As degeneracy exists in some variables (e.g., within-zone investment allocation, network operator capacity allocation, and unserved energy when price cap hits multiple price zones simultaneously), solutions may be non-unique. However, these multiple equilibria differ only in nodal quantities (except for unserved energy, which we treat ex-post, see Appendix II); aggregate installed capacity and zonal outcomes remain consistent across repeated runs.}

\begin{table}[ht]
\caption{Final ADMM residuals and iteration counts across market designs. Reported residuals are the maximum value over all zones. The convergence tolerance is $\epsilon=0.1$}
\label{tab:admm_convergence}
\centering
\footnotesize
\setlength{\tabcolsep}{4pt}
\begin{tabular}{lccccc}
\toprule
\textbf{Design} &
\textbf{Iterations} &
$\max r^{p}_{\mathrm{E}}$ &
$\max r^{d}_{\mathrm{E}}$ &
$\max r^{p}_{\mathrm{CM}}$ &
$\max r^{d}_{\mathrm{CM}}$ \\
\midrule
EOM-ref       & 1,504 & 0.012 & 0.100 & 0     & 0     \\
EOM-cap       &   462 & 0.017 & 0.100 & 0     & 0     \\
CM-NoCBP      &   498 & 0.047 & 0.099 & 0.031 & 0.077 \\
CM-ref        &   845 & 0.084 & 0.038 & 0.092 & 0.024 \\
CM-Implicit   & 2,223 & 0.027 & 0.100 & 0.017 & 0.029 \\
CM-NTC        &   830 & 0.038 & 0.097 & 0.057 & 0.099 \\
CM-FBMC       &   733 & 0.031 & 0.047 & 0.025 & 0.088 \\
\bottomrule
\end{tabular}
\end{table}

\section*{Appendix II}\label{sec:appendix_2}

\textcolor{R1}{
During joint-scarcity hours, defined as hours in which at least two zones experience $\ENSzt>0$, these zones face the common price cap, so market clearing may not uniquely determine the zonal allocation of ENS. We resolve this multiplicity ex-post using Eq.~\eqref{eq:curtailment}, where $\ENSzt+\PezT-\Pcoz$ is the shortfall corrected for the zone's net position in the capacity market $\Pcoz$. A positive net position indicates an importing zone.}
\begin{subequations} 
\label{eq:curtailment} 
\begin{align} 
\tilde{\ENSzt} &= s_t \cdot \max\big\{0,\; \ENSzt + \PezT - \Pcoz\big\}, 
\label{eq:curtailment_position}\\
s_t &= \frac{\sum_z \ENSzt}
{\sum_z \max\big\{0,\; \ENSzt + \PezT - \Pcoz\big\}},
\label{eq:curtailment_proportion}\\
\tilde{\PezT} &= \PezT + \ENSzt - \tilde{\ENSzt}.
\label{eq:curtailment_flow}
\end{align} 
\end{subequations}
\textcolor{R1}{
Since $\sum_z\PezT=\sum_z\Pcoz=0$, $s_t\le1$, and total ENS is conserved by construction. Surplus zones receive no curtailment, while deficit zones share total ENS in proportion to their shortfalls. If no zone is in surplus, $s_t=1$ and $\tilde{\PezT}=\Pcoz$. The reallocation affects only the zonal distribution of ENS and net positions; all other outcomes remain unchanged.}

%
%
\printbibliography

@article{Bublitz2019AMechanisms,
    title = {{A survey on electricity market design: Insights from theory and real-world implementations of capacity remuneration mechanisms}},
    year = {2019},
    journal = {Energy Economics},
    author = {Bublitz, Andreas and Keles, Dogan and Zimmermann, Florian and Fraunholz, Christoph and Fichtner, Wolf},
    month = {5},
    pages = {1059--1078},
    volume = {80},
    publisher = {North-Holland},
    doi = {10.1016/J.ENECO.2019.01.030},
    issn = {0140-9883}
}

@techreport{ACER2024ACER2024,
    title = {{ACER Decision on ERAA 2024}},
    year = {2024},
    author = {{ACER}}
}

@article{Hoschle2018AnMarkets,
    title = {{An ADMM-Based Method for Computing Risk-Averse Equilibrium in Capacity Markets}},
    year = {2018},
    journal = {IEEE Transactions on Power Systems},
    author = {Hoschle, Hanspeter and Le Cadre, Hlne and Smeers, Yves and Papavasiliou, Anthony and Belmans, Ronnie},
    number = {5},
    month = {9},
    pages = {4819--4830},
    volume = {33},
    publisher = {Institute of Electrical and Electronics Engineers Inc.},
    doi = {10.1109/TPWRS.2018.2807738},
    issn = {08858950}
}

@article{Lete2022AnPerspective,
    title = {{An analysis of zonal electricity pricing from a long-term perspective}},
    year = {2022},
    journal = {Energy Economics},
    author = {L{\'{e}}t{\'{e}}, Quentin and Smeers, Yves and Papavasiliou, Anthony},
    month = {3},
    pages = {105853},
    volume = {107},
    publisher = {North-Holland},
    url = {https://www.sciencedirect.com/science/article/pii/S0140988322000366},
    doi = {10.1016/J.ENECO.2022.105853},
    issn = {0140-9883}
}

@article{Mays2019AsymmetricMarkets,
    title = {{Asymmetric risk and fuel neutrality in electricity capacity markets}},
    year = {2019},
    journal = {Nature Energy 2019 4:11},
    author = {Mays, Jacob and Morton, David P. and O’Neill, Richard P.},
    number = {11},
    month = {10},
    pages = {948--956},
    volume = {4},
    publisher = {Nature Publishing Group},
    url = {https://www.nature.com/articles/s41560-019-0476-1},
    doi = {10.1038/s41560-019-0476-1},
    issn = {2058-7546}
}

@article{Jimenez2024CanAgent-based-models,
    title = {{Can an energy only market enable resource adequacy in a decarbonized power system? A co-simulation with two agent-based-models}},
    year = {2024},
    journal = {Applied Energy},
    author = {Jimenez, I. Sanchez and Rib{\'{o}}-P{\'{e}}rez, D. and Cvetkovic, M. and Kochems, J. and Schimeczek, C. and de Vries, L. J.},
    month = {4},
    pages = {122695},
    volume = {360},
    publisher = {Elsevier},
    doi = {10.1016/J.APENERGY.2024.122695},
    issn = {0306-2619}
}

@misc{SEMO2025CapacityMarket,
    title = {{Capacity Market}},
    year = {2025},
    booktitle = {https://www.sem-o.com/markets/capacity-market-overview{\#}:{\~{}}:text=The{\%}20Capacity{\%}20Market{\%}20is{\%}20a,security{\%}20of{\%}20supply{\%}20of{\%}20electricity.},
    author = {{SEMO}},
    url = {https://www.sem-o.com/markets/capacity-market-overview}
}

@misc{EliaGroup2025CapacityMechanism,
    title = {{Capacity Remuneration Mechanism}},
    year = {2025},
    author = {{Elia Group}},
    url = {https://www.elia.be/en/electricity-market-and-system/adequacy/capacity-remuneration-mechanism}
}

@misc{Adekola2026CodeSupplement,
    title = {{Code supplement}},
    year = {2026},
    author = {Adekola, Kamaldeen and Bruninx, Kenneth},
    url = {https://github.com/kamaladekola/cross_border-CM}
}

@article{Roques2019CountingMechanisms,
    title = {{Counting on the neighbours: challenges and practical approaches for cross-border participation in capacity mechanisms}},
    year = {2019},
    journal = {Oxford Review of Economic Policy},
    author = {Roques, Fabien},
    number = {2},
    month = {4},
    pages = {332--349},
    volume = {35},
    publisher = {Oxford Academic},
    url = {https://dx.doi.org/10.1093/oxrep/grz008},
    doi = {10.1093/OXREP/GRZ008},
    issn = {0266-903X}
}

@article{Bhagwat2017Cross-borderSystems,
    title = {{Cross-border effects of capacity mechanisms in interconnected power systems}},
    year = {2017},
    journal = {Utilities Policy},
    author = {Bhagwat, Pradyumna C. and Richstein, Jörn C. and Chappin, Emile J.L. and Iychettira, Kaveri K. and De Vries, Laurens J.},
    month = {6},
    pages = {33--47},
    volume = {46},
    publisher = {Pergamon},
    doi = {10.1016/J.JUP.2017.03.005},
    issn = {0957-1787}
}

@article{Meyer2015Cross-borderIntegration,
    title = {{Cross-border effects of capacity mechanisms: Do uncoordinated market design changes contradict the goals of the European market integration?}},
    year = {2015},
    journal = {Energy Economics},
    author = {Meyer, Roland and Gore, Olga},
    month = {9},
    pages = {9--20},
    volume = {51},
    publisher = {North-Holland},
    doi = {10.1016/J.ENECO.2015.06.011},
    issn = {0140-9883}
}

@techreport{Menegatti2024Cross-BorderExternalities,
    title = {{Cross-Border participation: A false hope for fixing capacity market externalities?}},
    year = {2024},
    author = {Menegatti, Emma and Meeus, Leonardo},
    url = {www.eui.eu},
    institution = {European University Institute Robert Schuman Centre for Advanced Studies Florence School of Regulation, Loyola de Palacio Chair}
}

@article{Boyd2011DistributedMultipliers,
    title = {{Distributed Optimization and Statistical Learning via the Alternating Direction Method of Multipliers}},
    year = {2011},
    journal = {Foundations and Trends{\textregistered} in Machine Learning},
    author = {Boyd, Stephen and Parikh, Neal and Chu, Eric and Peleato, Borja and Eckstein, Jonathan},
    number = {1},
    month = {7},
    pages = {1--122},
    volume = {3},
    doi = {10.1561/2200000016},
    issn = {1935-8237}
}

@techreport{FederalMinistryforEconomicAffairsandClimateActionBMWK2024ElectricitySystem,
    title = {{Electricity market design of the future Options for a secure, affordable and sustainable electricity system}},
    year = {2024},
    author = {{Federal Ministry for Economic Affairs and Climate Action (BMWK)}},
    month = {8},
    url = {www.bmwk.de},
    institution = {},
    address = {Berlin}
}

@misc{ENTSO-E2025ENTSO-EAccess.,
    title = {{ENTSO-E Transparency Platform (API access).}},
    year = {2025},
    author = {{ENTSO-E}},
    url = {https://transparency.entsoe.eu/}
}

@misc{ENTSO-E2025ERAAAssessment,
    title = {{ERAA - European Resource Adequacy Assessment}},
    year = {2025},
    author = {{ENTSO-E}},
    url = {https://www.entsoe.eu/eraa/}
}

@techreport{ENTSO-E2025EuropeanTransition,
    title = {{European Network of Transmission System Operators for Electricity The role of Capacity Mechanisms to enable a secure and competitive energy transition}},
    year = {2025},
    author = {{ENTSO-E}},
    institution = {ENTSO-E}
}

@article{DeVries2007GenerationJob,
    title = {{Generation adequacy: Helping the market do its job}},
    year = {2007},
    journal = {Utilities Policy},
    author = {De Vries, Laurens J.},
    number = {1},
    month = {3},
    pages = {20--35},
    volume = {15},
    publisher = {Pergamon},
    url = {https://www.sciencedirect.com/science/article/pii/S0957178706000488},
    doi = {10.1016/J.JUP.2006.08.001},
    issn = {0957-1787}
}

@article{Bucksteeg2019ImpactMarket,
    title = {{Impact of coordinated capacity mechanisms on the European power market}},
    year = {2019},
    journal = {Energy Journal},
    author = {Bucksteeg, Michael and Spiecker, Stephan and Weber, Christoph},
    number = {2},
    pages = {221--264},
    volume = {40},
    publisher = {International Association for Energy Economics},
    url = {/doi/pdf/10.5547/01956574.40.2.mbuc?download=true},
    doi = {10.5547/01956574.40.2.MBUC;PAGE:STRING:ARTICLE/CHAPTER},
    issn = {01956574}
}

@article{Hoschle2018InefficienciesMarket,
    title = {{Inefficiencies caused by non-harmonized capacity mechanisms in an interconnected electricity market}},
    year = {2018},
    journal = {Sustainable Energy, Grids and Networks},
    author = {H{\"{o}}schle, H. and Le Cadre, H. and Belmans, R.},
    month = {3},
    pages = {29--41},
    volume = {13},
    publisher = {Elsevier},
    doi = {10.1016/J.SEGAN.2017.11.002},
    issn = {2352-4677}
}

@article{Mastropietro2015NationalNeighbours,
    title = {{National capacity mechanisms in the European internal energy market: Opening the doors to neighbours}},
    year = {2015},
    journal = {Energy Policy},
    author = {Mastropietro, Paolo and Rodilla, Pablo and Batlle, Carlos},
    number = {1},
    month = {7},
    pages = {38--47},
    volume = {82},
    publisher = {Elsevier},
    doi = {10.1016/J.ENPOL.2015.03.004},
    issn = {0301-4215}
}

@article{OfficeoftheEuropeanUnionL-2024RegulationRelevance.,
    title = {{Regulation (EU) 2024/1747 of the European Parliament and of the Council of 13 June 2024 amending Regulations (EU) 2019/942 and (EU) 2019/943 as regards improving the Union’s electricity market designText with EEA relevance.}},
    year = {2024},
    author = {Office of the European Union L-, Publications and Luxembourg, Luxembourg},
    url = {http://data.europa.eu/eli/reg/2024/1747/oj}
}

@techreport{ACER2024SecurityReport,
    title = {{Security of EU electricity supply 2024 Monitoring Report}},
    year = {2024},
    author = {{ACER}},
    url = {www.acer.europa.eu},
    institution = {ACER}
}

@unpublished{Menegatti2025ThreeEU,
    title = {{Three Steps to a Regional Capacity Market in the EU}},
    year = {2025},
    author = {Menegatti, Emma and Meeus, Leonardo},
    series = {RSC 2025/44},
    url = {www.eui.eu},
    institution = {European University Institute Robert Schuman Centre for Advanced Studies Florence School of Regulation, Loyola de Palacio Chair}
}

@article{Schonheit2021TowardTrading,
    title = {{Toward a fundamental understanding of flow-based market coupling for cross-border electricity trading}},
    year = {2021},
    journal = {Advances in Applied Energy},
    author = {Sch{\"{o}}nheit, David and Kenis, Michiel and Lorenz, Lisa and M{\"{o}}st, Dominik and Delarue, Erik and Bruninx, Kenneth},
    month = {5},
    pages = {100027},
    volume = {2},
    publisher = {Elsevier},
    doi = {10.1016/J.ADAPEN.2021.100027},
    issn = {2666-7924}
}

@article{Aravena2021TransmissionMarkets,
    title = {{Transmission capacity allocation in zonal electricity markets}},
    year = {2021},
    journal = {Operations Research},
    author = {Aravena, Ignacio and L{\'{e}}t{\'{e}}, Quentin and Papavasiliou, Anthony and Smeers, Yves},
    number = {4},
    month = {7},
    pages = {1240--1255},
    volume = {69},
    publisher = {INFORMS Inst.for Operations Res.and the Management Sciences},
    url = {http://pubsonline.informs.org.https://doi.org/10.1287/opre.2020.2082http://www.informs.orghttp://pubsonline.informs.org/journal/opreContact:aravenasolis1@llnl.gov,https://orcid.org/0000-0003-4837-3466},
    doi = {10.1287/OPRE.2020.2082/SUPPL{\_}FILE/OPRE.2020.2082.SM1.PDF},
    issn = {15265463}
}

@article{Conejo2023WhyPricing,
    title = {{Why Marginal Pricing?}},
    year = {2023},
    journal = {Journal of Modern Power Systems and Clean Energy},
    author = {Conejo, Antonio J.},
    number = {3},
    month = {5},
    pages = {693--697},
    volume = {11},
    publisher = {State Grid Electric Power Research Institute Nanjing Branch},
    doi = {10.35833/MPCE.2023.000064},
    issn = {21965420}
}

@article{Keppler2022WhyMarkets,
    title = {{Why the sustainable provision of low-carbon electricity needs hybrid markets}},
    year = {2022},
    journal = {Energy Policy},
    author = {Keppler, Jan Horst and Quemin, Simon and Saguan, Marcelo},
    month = {12},
    pages = {113273},
    volume = {171},
    publisher = {Elsevier},
    doi = {10.1016/J.ENPOL.2022.113273},
    issn = {0301-4215}
}

\end{document}